\documentclass[preprint,12pt,authoryear]{elsarticle}

\usepackage{amssymb}
\usepackage{amsthm}

\theoremstyle{definition}
\newtheorem{theorem}{Theorem}

\usepackage{float}
\usepackage{amsmath}
\usepackage{xcolor}
\usepackage{placeins}
\usepackage{hyperref}
\usepackage{bm}
\usepackage{dsfont}
\usepackage{multirow}
\usepackage{geometry}
\usepackage{rotating}
\usepackage{subcaption}
\usepackage{tikz}
\usetikzlibrary{calc}
\usepackage{graphicx}
\usepackage{pdflscape}
\usepackage{pgfplots}
\usepackage{algorithm}
\usepackage{algpseudocode}
\usepackage{eurosym}

\DeclareMathOperator*{\argmin}{\arg\!\min}

\definecolor{seabornblue}{HTML}{0173b2}
\definecolor{seaborngreen}{HTML}{029e73}

\emergencystretch=\maxdimen
\pgfplotsset{compat=1.18}

\journal{}

\begin{document}

\begin{frontmatter}

%% Title, authors and addresses

%% use the tnoteref command within \title for footnotes;
%% use the tnotetext command for theassociated footnote;
%% use the fnref command within \author or \affiliation for footnotes;
%% use the fntext command for theassociated footnote;
%% use the corref command within \author for corresponding author footnotes;
%% use the cortext command for theassociated footnote;
%% use the ead command for the email address,
%% and the form \ead[url] for the home page:
%% \title{Title\tnoteref{label1}}
%% \tnotetext[label1]{}
%% \author{Name\corref{cor1}\fnref{label2}}
%% \ead{email address}
%% \ead[url]{home page}
%% \fntext[label2]{}
%% \cortext[cor1]{}
%% \affiliation{organization={},
%%            addressline={}, 
%%            city={},
%%            postcode={}, 
%%            state={},
%%            country={}}
%% \fntext[label3]{}

\title{Enhancing reliability in prediction intervals using point forecasters: Heteroscedastic Quantile Regression and Width-Adaptive Conformal Inference}

%% use optional labels to link authors explicitly to addresses:
%% \author[label1,label2]{}
%% \affiliation[label1]{organization={},
%%             addressline={},
%%             city={},
%%             postcode={},
%%             state={},
%%             country={}}
%%
%% \affiliation[label2]{organization={},
%%             addressline={},
%%             city={},
%%             postcode={},
%%             state={},
%%             country={}}

\author[label1,label2]{Carlos Sebastián}

\affiliation[label1]{organization={Fortia Energía},
             addressline={Calle de Gregorio Benítez},
             city={Madrid},
             postcode={28043},
             country={Spain}}

\affiliation[label2]{organization={Universidad Politécnica de Madrid},
            city={Madrid},
            country={Spain}}
\ead{carlos.sebastian@alumnos.upm.es}
            
\author[label3,label4]{Carlos E. González-Guillén}

\affiliation[label3]{organization={Departamento de Matemática Aplicada a la Ingeniería Industrial, Escuela Técnica Superior de Ingenieros Industriales, Universidad Politécnica de Madrid},
            addressline={Calle de José Gutiérrez Abascal}, 
            city={Madrid},
            postcode={28006},
            country={Spain}}

\affiliation[label4]{organization={Instituto de Ciencias Matemáticas (CSIC-UAM-UC3M-UCM)},
            addressline={Calle Nicolás Cabrera}, 
            city={Madrid},
            postcode={28049},
            country={Spain}}
            
\author[label5]{Jesús Juan}

\affiliation[label5]{organization={Laboratorio de Estadística, Escuela Técnica Superior de Ingenieros Industriales, Universidad Politécnica de Madrid},
            addressline={Calle de José Gutiérrez Abascal}, 
            city={Madrid},
            postcode={28006},
            country={Spain}}

\begin{abstract}
Constructing prediction intervals for time series forecasting is challenging, particularly when practitioners rely solely on point forecasts. While previous research has focused on creating increasingly efficient intervals, we argue that standard measures alone are inadequate. Beyond efficiency, prediction intervals must adapt their width based on the difficulty of the prediction while preserving coverage regardless of complexity.
To address these issues, we propose combining Heteroscedastic Quantile Regression (HQR) with Width-
Adaptive Conformal Inference (WACI). This integrated
procedure guarantees theoretical coverage and enables
interval widths to vary with predictive uncertainty. We assess its performance using both a synthetic example and a real world Electricity Price Forecasting scenario. Our results show
that this combined approach meets or surpasses typical benchmarks for validity and efficiency, while also fulfilling important yet often overlooked practical requirements.
\end{abstract}


\end{frontmatter}

% \tableofcontents
% \newpage

\section{Context of the problem}\label{sec:context}

Probabilistic forecasting for time series is a key tool in fields such as finance, energy, and operations management, where understanding the range of potential future outcomes is as important as predicting the most likely value for optimal decision-making \citep{gneiting2014probabilistic}. Given a time series $y_1, y_2, \dots, y_T$, $y_t \in \mathbb{R}$, the objective is to forecast the next $h$ steps, $y_{T+1}, y_{T+2}, \dots, y_{T+h}$. Unlike point forecasts that provide a single estimated value, probabilistic forecasting aims to model the conditional density $\mathbb{P}\left( y_{T+i} \left \vert y_1, y_2, \dots, y_T, X_1, X_2, \dots, X_{T+i} \right. \right)$, for each of the next $h$ steps, where $X_t = \left( x_{t, 1}, x_{t, 2}, \dots, x_{t, k} \right)$ represents a vector of $k$ regressors available at each time step $t$.\\

The importance of probabilistic forecasting is particularly evident in high-stakes applications. For example, this can be seen in the electricity market through electricity price forecasting or renewable energy prediction \citep{zhang2014review, nowotarski2018recent}. In the first case, high price volatility requires the presence of risk-aware strategies. Underestimating future prices may result in missed opportunities for higher profits, while overestimating prices could lead to losses from overcommitment or uncompetitive bidding. Probabilistic forecasts allow market participants to incorporate these cost asymmetries into their strategies. In renewable energy forecasting, the inherent unpredictability of weather conditions creates a pressing need for probabilistic methods. The power output of renewable sources like wind turbines and solar panels depends heavily on stochastic variables such as wind speed or solar irradiance. Probabilistic forecasts enable grid operators to account for this variability by scheduling backup generation and maintaining reserve capacity. They also help renewable energy producers avoid penalties by providing a clearer picture of the likelihood of over or under-delivery in their market bids.\\

Let's assume that only one-step predictions are made ($h=1$) and that the conditional distribution to be modelled is unimodal. In this work, uncertainty is represented through prediction intervals, which provide a range within which future values are expected to fall with a specified probability. That is, given a miscoverage rate $\alpha \in \left(0, 1\right)$ a prediction interval $\widehat{C}_{\alpha}\left(X_{T+1}\right) = \left[ \widehat{l}_{\alpha}\left(X_{T+1}\right), \widehat{u}_{\alpha}\left(X_{T+1}\right) \right] \subseteq \mathbb{R}$ is build such that \begin{equation}\label{eq:marginal_coverage}
\mathbb{P}\left(y_{T+1} \in \widehat{C}_{\alpha}\left(X_{T+1}\right) \right) \geq 1 - \alpha.
\end{equation}

An interval is said to be valid when property (\ref{eq:marginal_coverage}) is satisfied, i.e. when its marginal coverage is greater than or equal to the target coverage determined by the user. However, coverage alone does not guarantee optimal intervals. When building a prediction interval, the most efficient valid interval possible is desired \citep{shafer2008tutorial}. The efficiency of an interval is related to its length. When two prediction intervals achieve the same coverage level, the one with shorter length is preferred. This preference stems from the fact that narrower intervals provide more precise and actionable information, reducing the uncertainty range while maintaining reliability. Achieving the specified level of coverage with minimal interval length is crucial for constructing intervals that are not only statistically valid but also practically useful. The interval length of a prediction interval $\widehat{C}_{\alpha}\left(X_{T+1}\right)$ is denoted by $\left\vert \widehat{C}_{\alpha}\left(X_{T+1}\right)\right\vert$.\\

While these are crucial properties, this work asserts that they alone are not sufficient for ensuring practical utility. In addition to these considerations, we propose two further essential properties that prediction intervals must satisfy to enhance their applicability in real-world decision-making.

\begin{enumerate}
    \item \textbf{Adaptivity and correlation with prediction difficulty}: prediction intervals should adapt to different levels of uncertainty present in the data. Specifically, the length of the intervals should be correlated with the difficulty of the prediction, such that shorter intervals are associated with easier-to-predict situations and longer intervals with more challenging ones. This ensures that the intervals effectively reflect the underlying uncertainty in the forecasting process.

    \item \textbf{Independence between coverage and interval length}: In practice, the assessment of a prediction interval's validity and efficiency requires a comprehensive evaluation over a period involving multiple prediction intervals. The empirical coverage is defined as $$ \frac{1}{N} \sum_{t=T}^{T+N} \mathds{1}\left( y_{t+1} \in  \widehat{C}_{\alpha}\left(X_{t+1}\right) \right)$$ where $\mathds{1}\left( \cdot \right)$ is the indicator function and $N \in \mathbb{N}$ is the number of predictions that have been made. The marginal coverage of $\widehat{C}_{\alpha}\left(X_{T+1}\right)$ is approximated by this quantity. That is, $$\widehat{\mathbb{P}}\left( y_{t} \in \widehat{C}_{\alpha}\left(X_{t}\right) \right) = \frac{1}{N} \sum_{t=T}^{T+N} \mathds{1}\left( y_{t+1} \in  \widehat{C}_{\alpha}\left(X_{t+1}\right) \right).$$ Let $\mathcal{I}_\rho$ be the set of indices such that the length of the interval associated with that index is within $\delta > 0$ of $\rho \in \mathbb{R}$, i.e, $$ \mathcal{I}_\rho = \left\lbrace t+1 \, : \,  \left \vert \widehat{C}_{\alpha}\left(X_{t+1}\right)  -  \rho \right \vert \leq \delta\,, \, \rho\in \mathbb{R} \right\rbrace. $$ Let $N_\rho$ be the number of elements of that set. For all $\rho$ such that $N_\rho \neq 0$, the desired property that we advocate is
    \begin{equation}\label{eq:independence_cov_il}
    \widehat{\mathbb{P}} \left( y_{t} \in \widehat{C}_{\alpha}\left(X_{t}\right)\left \vert \left \vert \widehat{C}_{\alpha}\left(X_{t}\right) \right \vert \right. \approx \rho  \right)\footnote{Here using $\left\vert \widehat{C}_{\alpha}\left(X_{t}\right) \right\vert \approx \rho$ means $\left \vert \widehat{C}_{\alpha}\left(X_{t+1}\right)  -  \rho \right \vert \leq \delta$, thus allowing a small difference $\delta$ around $\rho$. This way, you avoid requiring the interval length to match $\rho$ exactly, which can be restrictive or even infeasible depending on the data.} = \frac{1}{N_\rho} \sum_{i \, \in \, \mathcal{I}_\rho } \mathds{1}\left( y_{i} \in  \widehat{C}_{\alpha}\left(X_{i}\right) \right) = 1-\alpha
    \end{equation}
    for a specified miscoverage value $\alpha \in \left(0, 1 \right)$. This ensures that the coverage level is consistent and independent of the interval's length, maintaining its reliability across different prediction difficulties.
\end{enumerate}

While the first of these properties has been recognised in numerous papers as a desirable characteristic \citep{angelopoulos2021gentle}, to our knowledge there is only one paper \citep{feldman2021improving} that has dealt with the second and with a different perspective. In \cite{feldman2021improving}, independence between the coverage indicator and the interval length is pursued to enhance the conditional coverage of a quantile regression process. This approach is grounded in the observation that, for the true quantiles of the conditional distribution, the coverage indicator and the interval length are orthogonal. In this work this approximation is extended to the case of time series, but with the similar objective of making a better approximation of the real quantiles, thus avoiding biases in situations of low or high uncertainty.\\

Moreover, these properties are intrinsically linked to aleatoric and epistemic uncertainty. Adaptability to more or less predictable situations is directly tied to aleatoric uncertainty, as it ensures that prediction intervals accurately reflect variations in inherent randomness. For instance, periods of higher or lower volatility can be effectively differentiated. In contrast, the independence between coverage and interval length is associated with epistemic uncertainty, as it indicates the absence of biases linked to prediction difficulty. A method that produces noticeably different coverage levels depending on the interval length exhibits signs of inadequate modelling or insufficient data to capture the true data-generating process.\\

Finally, this work addresses a practical scenario commonly encountered in the industry, where only $M$ point forecasting models are available, and no additional information about the underlying data-generating process or the models themselves is accessible. Specifically, at time $T$, the only available information consists of the $M$ predictions $\bm{\widehat{y}}_{T+1} = \left(\widehat{y}_{T+1, 1},\, \widehat{y}_{T+1, 2},\, \dots,\, \widehat{y}_{T+1, M}\right)$ for $y_{T+1}$ along with their historical values. This situation is typical in contexts where organizations rely on external forecasting tools without detailed knowledge of their construction or assumptions. The primary goal of this work is to propose methods for generating reliable prediction intervals using only the outputs of these point forecasters, i.e, we want to build a prediction interval $\widehat{C}_{\alpha}(X_{T+1}) \equiv \widehat{C}_{\alpha}(\bm{\widehat{y}}_{T+1})$, at the same time that the resulting intervals exhibit the two key properties introduced earlier, which are often overlooked in the literature.  By addressing this limitation, the proposed approach ensures both theoretical soundness and practical applicability in industrial settings, making it a versatile solution for real-world forecasting challenges. To simplify the notation, in the rest of the paper the constructed interval will be denoted by $\widehat{C}_{\alpha, T+1}$, although it should be noted that the source of information in the construction comes from the different predictions.\\

The contributions of the paper are as follows:
\begin{enumerate}
    \item A quantile regression model is proposed, inspired by the philosophy of the Quantile Regression Averaging (QRA) model of \cite{nowotarski2015computing}, but with modifications so that there is an increasing relationship between the length of the interval and the difficulty of the prediction. Due to the particular use of the standard deviation of the point predictors, the model is called Heteroscedastic Quantile Regression (HQR).\\

    \item To provide theoretical coverage guarantees and to achieve uniformity of coverage regardless of the difficulty of the prediction, the Width-Adaptive Conformal Inference (WACI) method is proposed, which modifies the Adaptive Conformal Inference (ACI, \cite{gibbs2021adaptive}) method by solving the problems that the rest of the models in the literature may present in this regard.\\
\end{enumerate}

The combination of HQR with WACI ensures strong results in terms of both validity and efficiency while simultaneously fulfilling the two desired properties.\\

The rest of the paper is structured in the following way. Section \ref{sec:previous_work} discusses the different works related to uncertainty quantification in various forms and goes into detail on some of them (Sections \ref{sec:quantile_regression}, and \ref{sec:conformal_prediction}), as they are the basis of the different contributions of the paper. Section \ref{sec:our_proposal} details our proposal, distinguishing HQR (Section \ref{sec:hqr}) and WACI (Section \ref{sec:waci}), which is evaluated with a synthetic example presented in Section \ref{sec:synthetic_example} and with a real life electricity price forecasting example in Section \ref{sec:desc_epf}. Conclusions and future work finish the paper in Section \ref{sec:conclusions}.

\section{Prior work on probabilistic forecasting}\label{sec:previous_work}

\subsection{General overview}
Bayesian methods, by their very nature, are clear candidates for probabilistic prediction. Through Bayes' theorem, a posterior distribution can be obtained by updating beliefs as new information is obtained. Assuming a parametric model dependent on weights on the target variable, a distribution over these weights can be adopted. This is the approach followed in Bayesian neural networks \citep{neal2012bayesian}. One can also consider the Bayesian approach directly on the target variable in the variant known as evidential regression \citep{amini2020deep} or with a functional approach through Gaussian processes \citep{rasmussen2003gaussian}. However, Bayesian methods present problems such as the choice of the prior distribution or the computational complexity.\\

Assuming a specific distribution, one can try to estimate the distribution of $y_{T+1}$ based on the information known at time $T$. This is done by methods such as NGBoost \citep{duan2020ngboost}, GAMLSS \citep{stasinopoulos2008generalized} as well as distributional neural networks and mixture density networks \citep{bishop1994mixture}. But the constraint of selecting a particular distribution can be quite restrictive. Data behaviour often evolve over time, and a fixed distribution may fail to remain valid as these changes occur. Additionally, it is common to assume simple distributions (e.g., the normal distribution), which often fail to capture the complex characteristics exhibited by real-world data, such as skewness, heavy tails, or multimodality. Conversely, selecting an overly complex distribution can lead to issues such as overfitting or excessive computational demands.\\

From a non-parametric point of view, classical methods such as bootstrapping the residuals to generate prediction intervals can be applied \citep{efron1987better}. However, the generality of the method tends not to produce the most satisfactory results. The application of quantile regression \citep{koenker1978regression} is also very popular, either through a linear model or by extending the method to more complex approaches such as neural networks \citep{cannon2011quantile}. Quantile regression is explained in more detail in Section \ref{sec:quantile_regression}\\

All these methods can be easily extended to time series problems (for example by considering autoregressive effects, which is common practice) but none of them can assure the marginal coverage needed to provide valid prediction intervals. The Conformal Prediction framework \citep{vovk2005algorithmic} ensures such marginal coverage in finite samples by assuming exchangeability between observations and without any assumptions about the probability distribution. In fact, Conformalized Quantile Regression (CQR) \citep{romano2019conformalized} extends quantile regression by providing the property of validity under exchangeability, while trying to fit the heteroscedasticity properties encountered in the data. Indeed, the properties that we want to be found in the intervals to be constructed are closely tied to achieving conditional coverage, a topic of significant interest in the field of conformal predictions \citep{romano2019conformalized, sesia2021conformal, chernozhukov2021distributional, han2022split, izbicki2022cd}. However, the existing body of work on conditional coverage does not fully align with the framework considered here, as it typically assumes a static data distribution over time, a condition that is not met in our context. As the exchangeability property is very demanding in time series, a large branch of research has focused on maintaining the good properties of the conformal predictors without assuming it. See for example \citep{gibbs2021adaptive, zaffran2022adaptive, bhatnagar2023improved, auer2023conformal, gibbs2022conformal}. More details related with Conformal Prediction can be found in Section \ref{sec:conformal_prediction} \\

Regarding the context of the problem at hand, where only different predictors of the event to be forecasted are known, most methodologies can, in principle, be adapted by treating these predictors as explanatory variables. However, to our knowledge, there is only one work that has approached it in such a way: the Quantile Regression Averaging (QRA) model proposed by \cite{nowotarski2015computing}. Additionally, there are methods designed to combine predictors of the mean to enhance point forecasts, which have been adapted for probabilistic forecasting in an online setting, such as the approach proposed by \cite{gaillard2016additive} using the algorithm in \cite{gaillard2014second}. This paper will focus on QRA, which will be examined in greater detail in Section \ref{sec:QRA}.\\

\subsection{Quantile regression}\label{sec:quantile_regression}

Let $F_{T+1}$ be the cumulative distribution function of the random variable $Y_{T+1}$. For a given probability level $\beta \in \left(0, 1\right)$, the quantile $\beta$ of $Y_{T+1}$ is defined as $$q_{\beta}\left(Y_{T+1}\right)= \inf \left\lbrace y \in \mathbb{R} \, : \, F_{T+1} \left(y\right) \geq \beta \right\rbrace.$$

When constructing prediction intervals, prioritizing their validity is crucial. Valid intervals are essential for making informed decisions and managing risk, as they guarantee the desired level of coverage. If the intervals do not align with the expected confidence level, their reliability is compromised, leading to a loss of trust in the predictions and diminishing their value as decision-making tools. Therefore, the estimation of quantiles is a logical approximation to the problem. Let $\alpha \in (0, 1)$ be the target miscoverage value and $\xi \in (0, \alpha)$. If $C_{\alpha, T+1} = \left[ l_{\alpha, T+1}, u_{\alpha, T+1}\right] = \left[ q_{\xi}(Y_{T+1}), q_{1-\alpha+\xi}(Y_{T+1})\right]$ is considered, then $$\mathbb{P}\left( y_{T+1} \in C_{\alpha, T+1}\right) \geq 1-\alpha, $$ where $y_{T+1}$ is the actual observed value at time $T+1$.\\

The typical choice for $\xi$ is $\xi = \frac{\alpha}{2}$ and this is the approach that will be followed in this paper. However, it should be noted that if the smallest intervals are desired, this option is not necessarily optimal. For example, when the distribution of $Y_{T+1}$ is asymmetric, other choices of $\xi$ can provide shorter intervals without sacrificing coverage.\\

Let $\bm{y} = \left(y_1, y_2, \dots, y_n \right)$ be the vector of $n$ observations of the variable of interest and $\bm{x}_i = (x_{i, 1}, x_{i, 2}, \dots, x_{i, m})$ be the vector of $m$ explanatory features for the observation $i$ for all $i \in \left\lbrace 1, 2, \dots, n \right\rbrace$. Let $y_i \sim Y$ and $\bm{x}_i \sim X$ for all $i=1, 2, \dots, n$. Let $Y \vert X $ be distributed as $F$. Let's consider the model $$q_\beta(Y\vert \bm{x}_i) = \lambda_0(\beta) + \lambda_1(\beta) x_{i, 1} + \lambda_2(\beta) x_{i, 2} + \dots + \lambda_m(\beta) x_{i, m} + \varepsilon_i(\beta); \; \mathbb{E}\left[ \varepsilon_i(\beta) \right] = 0$$ where $\bm{\lambda}\left(\beta\right) \equiv \bm{\lambda} = \left(\lambda_0\left(\beta\right), \lambda_1\left(\beta\right), \lambda_2\left(\beta\right), \dots, \lambda_m\left(\beta\right) \right)$ are the parameters of the model and $\varepsilon_i(\beta)$ represents noise.\\ 

Just as the mean squared error serves as the loss function optimized to estimate the conditional mean as a point estimator, the conditional quantile is estimated by minimizing the pinball loss function \citep{koenker1978regression}:
\begin{equation*}\label{eq:pinball}
    \ell_{\beta}(y_i, \widehat{y}_i) = \beta\vert y_i - \widehat{y}_i \vert \mathds{1}\left\lbrace y_i - \widehat{y}_i \geq 0 \right\rbrace + (1-\beta)\vert y_i - \widehat{y}_i \vert \mathds{1}\left\lbrace y_i - \widehat{y}_i \leq 0 \right\rbrace
\end{equation*}

The parameters $\bm{\lambda}$ are estimated as $$\widehat{\bm{\lambda}} = \min_{\bm{\lambda}} \left\lbrace \sum_{i=1}^n \ell_\beta(y_i, \bm{\lambda}^{\text{T}} \bm{x}_i)\ \right\rbrace$$ and inference about a new observation $n+1$ is done through $$\widehat{q}_{\beta}(Y \vert \bm{x}_{n+1}) = \widehat{\lambda}_0\left(\beta\right) + \widehat{\lambda}_1\left(\beta\right) x_{n+1, 1} + \widehat{\lambda}_2\left(\beta\right) x_{n+1, 2} + \dots + \widehat{\lambda}_m\left(\beta\right) x_{n+1, m}$$

\subsubsection{Quantile Regression Averaging}\label{sec:QRA}

The core idea in \cite{nowotarski2015computing} is to estimate the quantiles by treating the individual point forecasts as independent variables. While the original work does not explicitly restrict itself to using only this information, the model presented there is well-suited to the current problem, which is to produce prediction intervals relying solely on different point predictors.\\ 

Although the model is presented in the context of Day-Ahead electricity price forecasting, it is perfectly generalizable to any regression problem. In particular, the model proposed for the quantile $\beta$ at time $t$ is 
\begin{equation}\label{eq:qra}
    q_{\beta}(Y_t \vert \bm{y}_{t}) = \lambda_{0}(\beta) + \lambda_{1}(\beta)  \widehat{y}_{t, 1} + \lambda_{2}(\beta)  \widehat{y}_{t, 2} + \dots + \lambda_{m}(\beta)  \widehat{y}_{t, M} + \varepsilon_t(\beta) ; \; \mathbb{E}\left[ \varepsilon_t(\beta)  \right] = 0,
\end{equation}
where $Y_t$ is the random variable associated to instant $t$ and $ \bm{y}_{t} = \left(\widehat{y}_{t, 1}, \widehat{y}_{t, 2}, \dots, \widehat{y}_{t, M}\right)$ are predictions of the mean from $M$ different models for the same time instant $t$. In particular, one-step ahead predictions would be obtained by:

\begin{equation}\label{eq:qra_inference}
\widehat{q}_{\beta}(Y_{T+1} \vert \bm{y}_{T+1}) = \widehat{\lambda}_{0}(\beta) + \widehat{\lambda}_{1}(\beta)  \widehat{y}_{T+1, 1} + \widehat{\lambda}_{2}(\beta)  \widehat{y}_{T+1, 2} + \dots + \widehat{\lambda}_{m}(\beta)  \widehat{y}_{T+1, M}.
\end{equation}

\begin{figure}[H]
    \centering
    \includegraphics[scale=0.65]{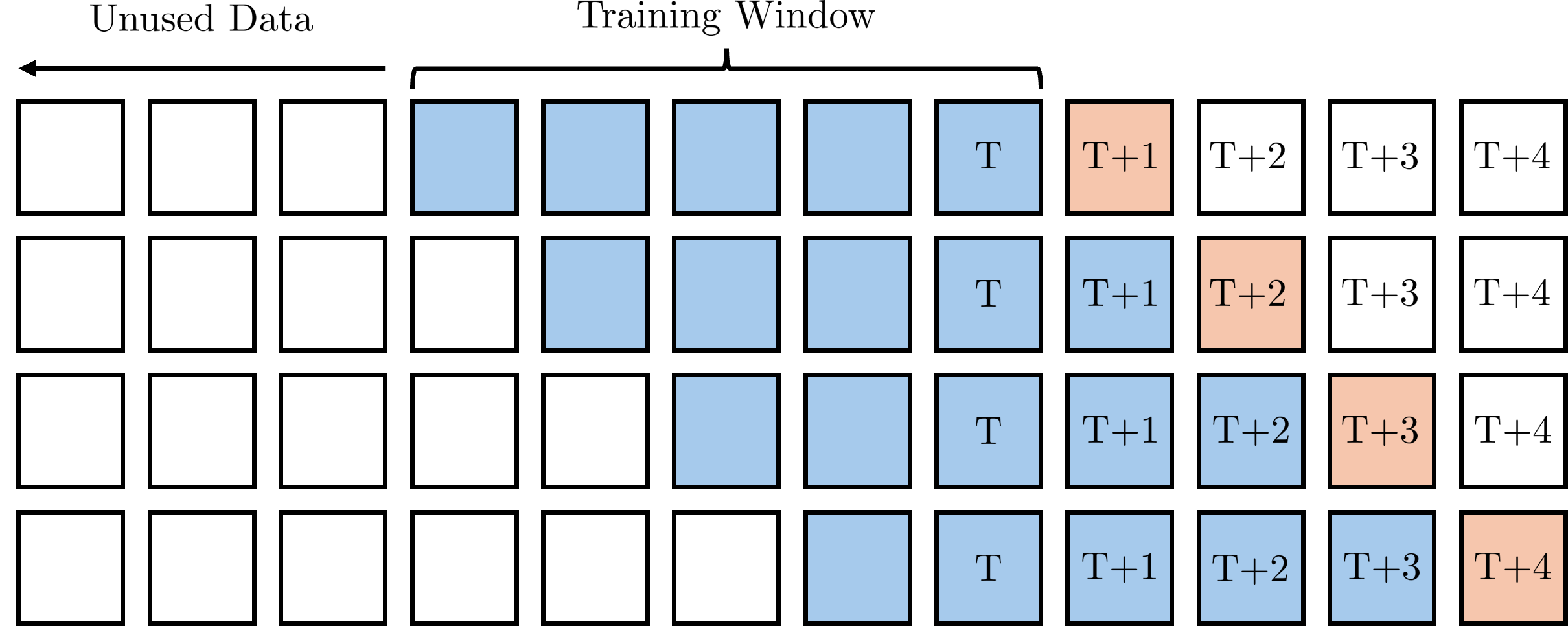}
    \caption{Rolling window mechanism with size equal to 5 time steps. To predict the next time step, only the data from the previous 5 time steps is used to estimate the model parameters.}
    \label{fig:rolling_window}
\end{figure}

To fit the models for time series problems, the use of a rolling window approach is proposed. Thus, the model described in equation (\ref{eq:qra}) and (\ref{eq:qra_inference}) would be the particular model for one window. For another window, another estimation of the model parameters would be obtained. Figure \ref{fig:rolling_window} describes the process of a rolling window methodology. The window size in this procedure is chosen empirically.\\

Although quantile regression procedures based on the pinball loss produce asymptotically consistent estimators \citep{koenker1978regression}, over a finite amount of data there is no theoretical guarantee of obtaining the desired marginal coverage. This is where the Conformal Prediction framework adds value.\\

\subsection{Conformal Prediction}\label{sec:conformal_prediction}

Conformal predictions were introduced in \cite{vovk2005algorithmic} to build prediction intervals (in the regression framework) that are valid with a finite number of data, without assumptions, except exchangeability, about any kind of distribution. It operates as a post-processing phase within an existing prediction pipeline, enabling the construction of valid prediction intervals without requiring any modifications to the existing forecasting process. Its fundamental base assures that intervals are valid regardless of the quality of the initial predictions, although their efficiency remains influenced by the accuracy of the preceding forecasting phase.\\

Although the original approach, commonly referred to as Full Conformal Prediction, is not computationally feasible on a large scale, the approach known as Split Conformal Prediction (SCP, \cite{lei2018distribution, papadopoulos2002inductive}) solves such problems making its use more appealing in a multitude of situations. This paper only focuses on the second approach.\\

Suppose we have $n$ points $(\bm{x}_i, y_i) \in \mathbb{R}^m \times \mathbb{R}, \, i = 1, \dots, n$ and we are interested in providing a prediction interval for the next observation $y_{n+1}$ for which $\bm{x}_{n+1}$ is known. Conformalization in its simplest form consists in making a correction to a prediction of the mean. Let $\widehat{\mu}(\cdot)$ be that predictor. The steps to perform its conformalization for an objective miscoverage of $\alpha$ are described in Algorithm \ref{alg:split_conformal}.

\begin{algorithm}
\caption{Conformalized Mean Regression trough SCP}\label{alg:split_conformal}
\begin{algorithmic}[1]
\Require $\{(\bm{x}_i, y_i)\}_{i=1}^n$, significance level $\alpha$, regression algorithm $\widehat{\mu}$
\State Randomly split the $n$ known points into two disjoint sets: training $\text{Tr}$ and calibration $\text{Cal}$.
\State Train the regression algorithm $\widehat{\mu}$ using the data from the training set $\text{Tr}$.
\State Compute the conformity scores for the calibration set $\text{Cal}$ using the absolute error:
$$\mathcal{S} = \mathcal{S}_{\text{Cal}} \cup \{+\infty\},$$
where 
$$\mathcal{S}_{\text{Cal}} = \{\vert y_i - \widehat{\mu}(\bm{x}_i) \vert \, : \, i \in \text{Cal}\}.$$
\State Compute the $(1-\alpha)$ quantile of the conformity scores, denoted as $Q_{1-\alpha}(\mathcal{S})$.
\State Construct the conformal prediction interval for observation $n+1$ as:
$$\widehat{C}_{\alpha, n+1} = \left[ \widehat{\mu}(\bm{x}_{n+1}) - Q_{1-\alpha}(\mathcal{S}), \; \widehat{\mu}(\bm{x}_{n+1}) + Q_{1-\alpha}(\mathcal{S}) \right].$$
\Ensure $\widehat{C}_{\alpha, n+1}$ prediction interval of level $1-\alpha$ for the observation $n+1$.
\end{algorithmic}
\end{algorithm}

\begin{theorem}[\cite{lei2018distribution}]
    Let $(\bm{x}_i, y_i)_{i=1}^{n+1}$ be exchangeable. The process of conformalizing a conditional mean predictor as described in Algorithm \ref{alg:split_conformal} produces a prediction interval for the observation $n+1$, $\widehat{C}_{\alpha, n+1}$,  such that $$ \mathbb{P}\left( y_{n+1} \in \widehat{C}_{\alpha, n+1} \right) \geq 1 - \alpha.$$
    If, in addition, the scores $\mathcal{S}_{\text{Cal}}$ have a continuous joint distribution, then: $$ \mathbb{P}\left( y_{n+1} \in \widehat{C}_{\alpha, n+1} \right) \leq 1 - \alpha + \dfrac{1}{\# \text{Cal}+1}.$$
\end{theorem}

\subsubsection{Conformalized Quantile Regression (CQR)}\label{sec:cqr}

While this methodology is useful, its simplicity does not take into account the possible heteroscedasticity depending on the covariates. That is, a stronger property that would be desirable is conditional coverage:
 $$ \mathbb{P}\left( y_{n+1} \in C_{\alpha, n+1} \vert \bm{x}_{n+1} = \bm{x} \right) \geq 1 - \alpha \quad \forall \, \bm{x} \in \mathbb{R}^m.$$

\begin{figure}[h]
     \centering
     \includegraphics[width=\linewidth]{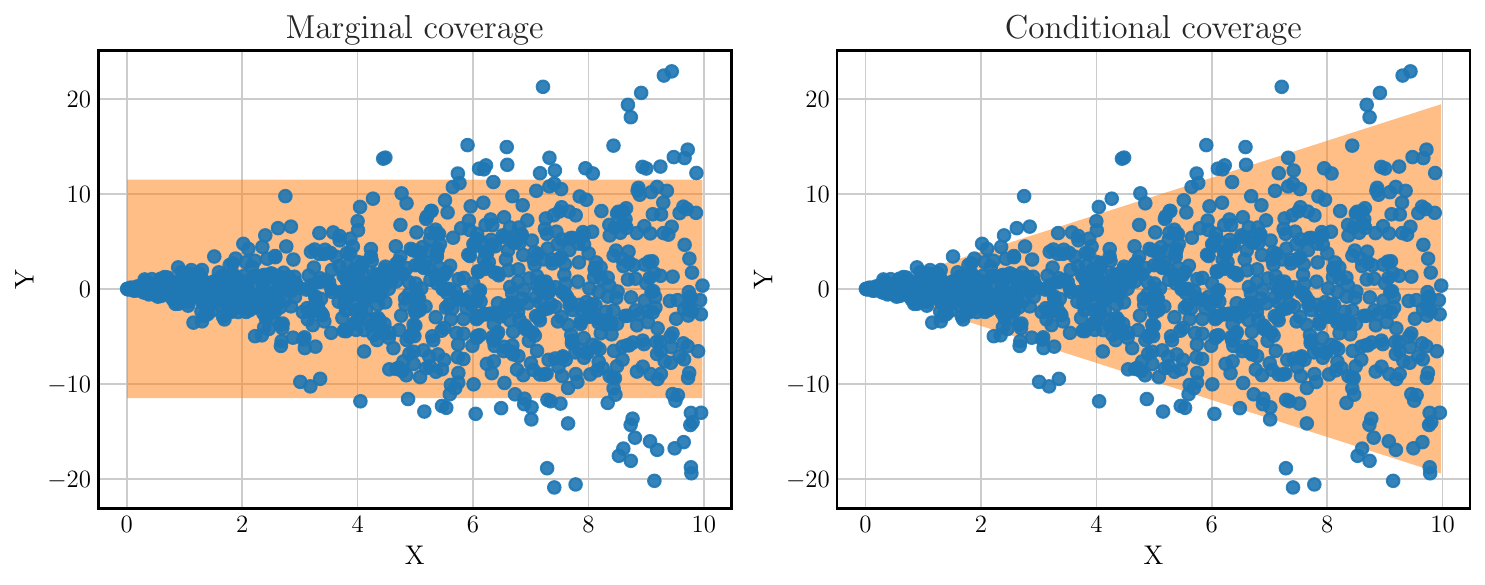}
     \caption{Difference between marginal coverage and conditional coverage in a toy dataset.}
     \label{fig:marginal_vs_conditional}
 \end{figure}

 In Figure \ref{fig:marginal_vs_conditional} the difference between marginal and conditional coverage is appreciated. Notice how in the case of conditional coverage, the prediction intervals are adjusted to the heteroscedasticity of the data as a function of $X$.\\

For any distribution-free approximation, not just Conformal Prediction, enforcing this property would require the intervals to be uninformative. That is, achieving this property in a practical way and without assuming any distribution is not possible \citep{vovk2012conditional, lei2014distribution}. Therefore, a variety of works have been developed to approximate it as best as possible. The most popular of these is probably the Conformalized Quantile Regression (CQR) procedure proposed in \cite{romano2019conformalized}. CQR follows the conformal methodology to correct the coverage obtained by estimating the quantiles through a quantile regression procedure. Since a quantile regression procedure cannot guarantee the desired coverage level in finite samples, CQR adjusts the interval bounds generated by this procedure. The intervals are enlarged or reduced if the empirical marginal coverage in a calibration set is found to be smaller or larger than the target level, respectively. Although there is no theoretical result related to conditional coverage, as the correction is performed on these estimated conditional quantiles, it is expected that heteroscedasticity is captured with much better quality than with the traditional conformal approach.\\

The CQR methodology for an objective miscoverage rate of $\alpha$ is describe in  Algorithm \ref{alg:cqr}.

\begin{algorithm}
\caption{Conformalized Quantile Regression}\label{alg:cqr}
\begin{algorithmic}[1]
\Require $\{(\bm{x}_i, y_i)\}_{i=1}^n$, significance level $\alpha$, quantile regression algorithm $\mathcal{A}$
\State Randomly split the $n$ known points into two disjoint sets: training $\text{Tr}$ and calibration $\text{Cal}$.
\State Train the quantile regression algorithm $\mathcal{A}$ using the data from the training set $\text{Tr}$ and obtain a first approximation of $l_{\alpha, i}$ and $u_{\alpha, i},\, \widehat{l}_{\alpha, i}$ and $\widehat{u}_{\alpha, i}, \, i \in \text{Cal}\cup\{n+1\}$
\State Compute the conformity scores $\mathcal{S}$
$$ \mathcal{S} = \left\lbrace \mathcal{S}_i \, : \, i \in \text{Cal} \right\rbrace \cup \left\lbrace +\infty \right\rbrace$$ where $\mathcal{S}_i = \max \left\lbrace y_i - \widehat{u}_{\alpha,i}, \widehat{l}_{\alpha,i} - y_i \right\rbrace$
\State Compute the $(1-\alpha)$ quantile of the conformity scores, denoted as $Q_{1-\alpha}(\mathcal{S})$.
\State Construct the conformal prediction interval for observation $n+1$ as:
$$\widehat{C}_{\alpha, n+1} = \left[ \widehat{l}_{\alpha, n+1} - Q_{1-\alpha}(\mathcal{S}), \widehat{u}_{\alpha, n+1} + Q_{1-\alpha}(\mathcal{S}) \right] $$
\Ensure $\widehat{C}_{\alpha, n+1}$ prediction interval of level $1-\alpha$ for the observation $n+1$.
\end{algorithmic}
\end{algorithm}

\begin{theorem}[\cite{romano2019conformalized}]
Let $\left(\bm{x}_i, y_i \right) _{i = 1}^{n+1}$ be exchangeable. Applying CQR $\left( \bm{x}_i, y_i \right) _{i = 1}^n$ produces a prediction interval $\widehat{C}_{\alpha, n+1}$ such that:
\begin{equation*}
\mathbb{P}\left(y_{n+1} \in \widehat{C}_{\alpha, n+1}\right) \geq 1-\alpha.
\end{equation*} 
Moreover, if the conformity scores $\{S_i\}_{i \in \text{Cal}}\cup \{+\infty\}$ are almost surely distinct, then the prediction interval is nearly perfectly calibrated:
\begin{equation*}
\mathbb{P}\left(y_{n+1} \in \widehat{C}_{\alpha, n+1}\right) \leq 1-\alpha +\dfrac{1}{\#\text{Cal}+1}.
\end{equation*} 
\end{theorem}

Given that the computational complexity is minimal once the quantile regression model is trained, and valid intervals are ensured on finite and exchangeable samples, CQR is considered the standard approach in Conformal Prediction for regression tasks. Therefore, it will serve as the foundation for the conformalizations performed through ACI and WACI (Sections \ref{sec:aci} and \ref{sec:waci}). For more details of the CQR algorithm we refer to \cite{romano2019conformalized}.\\

\subsubsection{Adaptive Conformal Inference (ACI)}\label{sec:aci}

CQR or any other conformal algorithm following the presented scheme (Algorithm \ref{alg:cqr}) depends on the condition of exchangeability among observations. In time series, which are the problems we are interested in, this condition is not fulfilled. Removing the condition of exchangeability while maintaining the validity property of the intervals has been one of the primary research objectives in the field. One such work is the Adaptive Conformal Inference (ACI) method proposed by \cite{gibbs2021adaptive}.\\

The application of ACI over the CQR procedure with $\alpha^*$ as the objective miscoverage rate looks as follows. Let $\alpha_1 = \alpha^*, \text{err}_1 = 0 $ and $ \gamma > 0$.
\[
\begin{cases}
    \alpha_{t+1} &= \alpha_t + \gamma(\alpha^*-\text{err}_t) \\
    \text{err}_t &= \begin{cases}
        1 & \text{if } y_t \not\in \widehat{C}_{\alpha^*, t} \\
        0 & \text{otherwise}
    \end{cases} \\
    \widehat{C}_{\alpha^*, t+1} &= \left[ \widehat{l}_{\alpha^*, t+1} - Q_{1-\alpha_{t+1}}(\mathcal{S}), \widehat{u}_{\alpha^*, t+1} + Q_{1-\alpha_{t+1}}(\mathcal{S}) \right] \\
\end{cases}
\]

It is a CQR procedure where the quantile used to make the correction is not necessarily that of the target coverage. It is taken adaptive depending on whether too large or too small intervals are being considered. The speed of adaptation is determined by the parameter $\gamma$. The following result can be derived:\\

\begin{theorem}[\cite{gibbs2021adaptive}]
    With probability one it follows that for all $T \in \mathbb{N}$, $$\left\vert \frac{1}{T}\sum_{t=1}^T \text{err}_t - \alpha^* \right\vert \leq \frac{\max \left\lbrace \alpha_1, 1-\alpha_1 \right\rbrace + \gamma}{T \gamma}.$$
In particular, $$\lim_{T \rightarrow \infty} \frac{1}{T} \sum_{t=1}^T \text{err}_t = \alpha^*.$$ In other words, there is asymptotic marginal coverage.
\end{theorem}

For more details on the ACI algorithm we refer to \cite{gibbs2021adaptive} and \cite{zaffran2022adaptive}.\\

\section{Our proposal}\label{sec:our_proposal}

Previous research has focused on the issue of providing valid and efficient prediction intervals for individual values. A comprehensive analysis of the coverage of prediction bands, as well as of the interval lengths associated, based on this individual approach reveals significant issues which were described in Section \ref{sec:context}.The first is that the length of the intervals varies depending on the difficulty of the observation to be predicted. That is, there should be an increasing relationship between the error of the point forecasting model and the length of the proposed interval.\\

One of the main works leveraging point predictors to build prediction intervals, as in the context of this study, is the QRA method described in Section \ref{sec:QRA}. However, its approach does not account for the desired property of adapting to heteroscedasticity. In this paper, we propose a model inspired by QRA, but explicitly designed to incorporate the heteroscedasticity. This design enables our approach to simultaneously capture the aleatoric uncertainty associated with the event and the epistemic uncertainty of the predictors, ensuring that the estimated uncertainty reflects the underlying complexity of the prediction task.\\

\subsection{Heteroscedastic Quantile Regression (HQR)}\label{sec:hqr}

The QRA model expresses the quantile of interest as a linear combination of point predictors of the mean. The effectiveness shown by this model manifests that the information given by different predictors of the event of interest provides information when quantifying the associated uncertainty.\\

\begin{figure}[h]
    \centering
    \hspace*{-2.5cm}
    \includegraphics[scale=0.6]{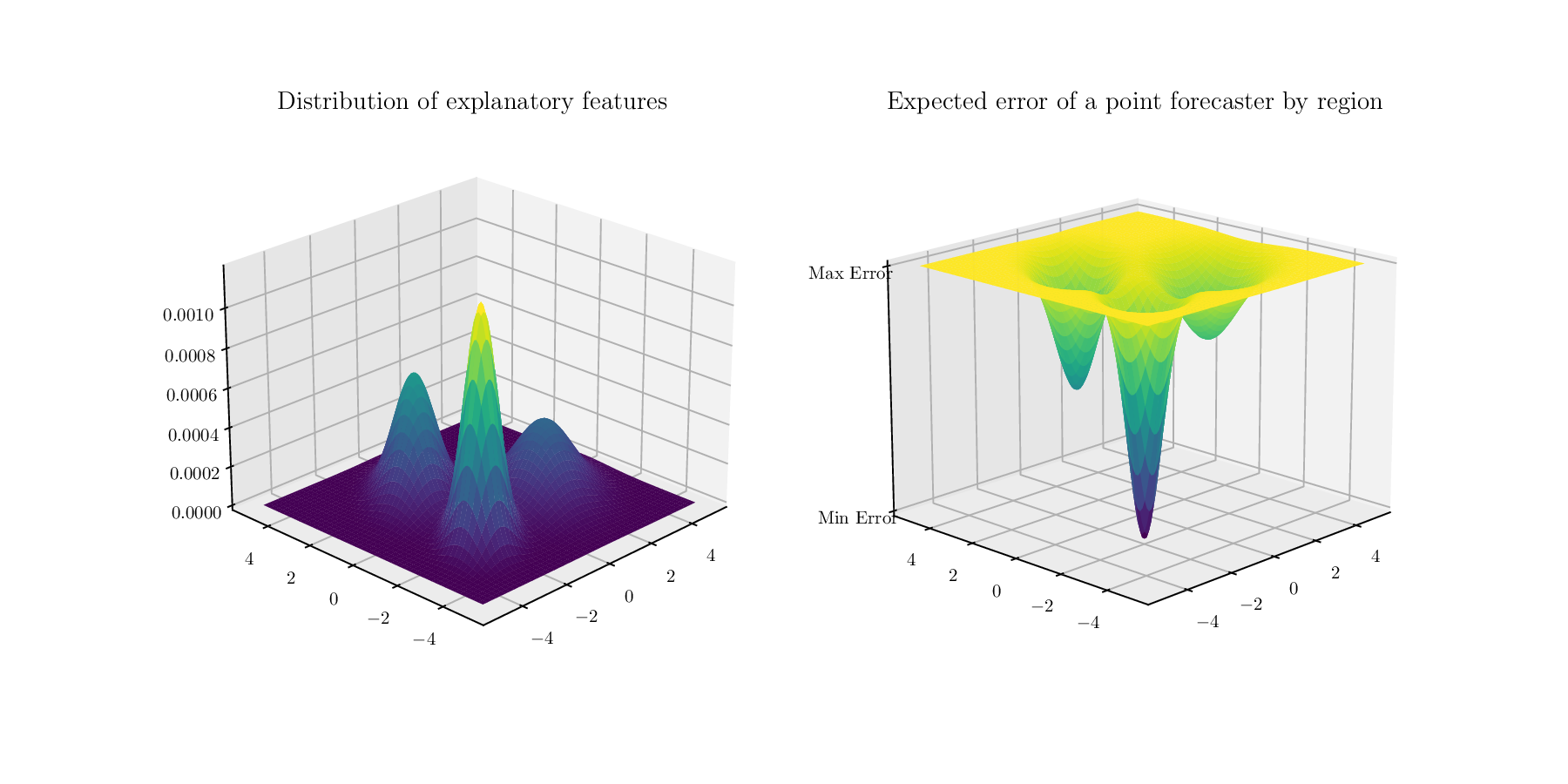}
    \caption{The joint distribution of two explanatory features is shown on the left. On the right, the expected error for a predictive model is plotted as a function of the two features. One would expect to have a higher error in the unexplored areas of the space, while a lower error would be expected in the very common areas. The plot is for guidance as the model could have good extrapolation properties in some situations.}
    \label{fig:expected_error}
\end{figure}

It is clear that having different predictors of the expected value can provide information on the safety of the prediction: in very common situations for the model, i.e., in areas where the space of regressor variables is highly explored, all forecasters are likely to obtain very similar predictions. However, in the more unfamiliar situations, which generally correspond to unexplored areas where models have to extrapolate, the forecasts start to differ, and, in particular, the error of the models in such cases is generally larger (Figure \ref{fig:expected_error}). This reflects higher epistemic uncertainty, as it stems from a lack of knowledge or a lack of data in some of these regions. Additionally, situations of high aleatoric uncertainty, where inherent randomness in the data-generating process dominates, can also lead to divergence in the predictions of the mean. In these cases, even with well-trained models, the variability in the predictions reflects the fundamental unpredictability of the process. In other words, both epistemic and aleatoric uncertainty contribute to the observed differences in the forecasts, which is a factor that should be taken into account when building prediction intervals. \\

A good indicator of the level of exploration of the explanatory features space is a dispersion measure of the prediction of the different models. Suppose, following the model of \cite{nowotarski2015computing}, that for each time $t$ we have $M$ different forecasts $\widehat{y}_{t, 1}, \widehat{y}_{t, 2}, \dots, \widehat{y}_{t, M}$ intended to predict $y_t$. Thus, denoting by $\overline{\widehat{y}}_{t} = \dfrac{1}{M} \displaystyle \sum_{i=1}^M \widehat{y}_{t, i}$ and $s_{\widehat{y}_{t}}^2 = \dfrac{1}{M} \displaystyle \sum_{i=1}^M \left(\widehat{y}_{t, i} - \overline{\widehat{y}}_{t} \right)^2$ the following quantile regression model is proposed:
\begin{equation}\label{eq:cfqra}
    q_{\beta}(Y_t \vert \bm{\widehat{y}}_t) = \lambda_{0}(\beta) + \lambda_{1}(\beta) \overline{\widehat{y}}_{t} + \lambda_{2}(\beta)  s_{\widehat{y}_{t}} + \varepsilon_t(\beta), \; \mathbb{E}\left[\varepsilon_t(\beta) \right] = 0,
\end{equation}
where the parameters are obtained by minimizing the pinball loss (equation (\ref{eq:pinball})) and vary over time using a rolling window procedure in the same way as in equation (\ref{eq:qra}). In particular, for the $\frac{\alpha}{2}$ and $1-\frac{\alpha}{2}$ quantiles of interest, we have:
$$
\begin{cases}
\widehat{q}_{\frac{\alpha}{2}}(Y_{T+1} \vert \bm{\widehat{y}}_{T+1}) &= \widehat{\lambda}_{0}(\frac{\alpha}{2}) + \widehat{\lambda}_{1}(\frac{\alpha}{2})\overline{\widehat{y}}_{T+1} + \widehat{\lambda}_{2}(\frac{\alpha}{2})s_{\widehat{y}_{T+1}}\\
\widehat{q}_{1-\frac{\alpha}{2}}(Y_{T+1} \vert \bm{\widehat{y}}_{T+1}) &= \widehat{\lambda}_{0}(1-\frac{\alpha}{2}) + \widehat{\lambda}_{1}(1-\frac{\alpha}{2})\overline{\widehat{y}}_{T+1} + \widehat{\lambda}_{2}(1-\frac{\alpha}{2})s_{\widehat{y}_{T+1}}\\
\end{cases}
$$

Intuitively, we would expect high values of the $\lambda_{2}(\beta)$ parameter for quantiles further away from the median with a positive sign for quantiles greater than 0.5 and a negative sign for quantiles less than 0.5. Similarly, smaller values of $\lambda_{2}(\beta)$ would be found for quantiles close to the median. If this behaviour occurs, then we would have the relationship between the length of the interval and the error that we are looking for (\ref{ap:behaviour_cfqra}).\\

Note that equation (\ref{eq:cfqra}) is actually an extension of the QRA model defined in (\ref{eq:qra}). In the case of the QRA model, what is being done is to estimate the mean through a weighted average, which results in different values of the coefficients $\lambda_1, \dots, \lambda_M$. That is, the QRA model is a model of the type $$q_{\beta}(Y_t \vert \bm{\widehat{y}}_t) = \lambda_{0}(\beta) + \lambda_{1}(\beta) \overline{\widehat{y}}_{t} + \varepsilon_t(\beta), \; \mathbb{E}\left[\varepsilon_t(\beta) \right] = 0,$$ where the estimation $\overline{\widehat{y}}_{t}$ is not done with equal weights. In that sense, we are extending the model with a further component that refers to a first assessment of the level of uncertainty that exists. Because this extension is directly related to the heteroscedasticity of the predictions, the model has been named Heteroscedastic Quantile Regression (HQR).\\

To evaluate the significance of mean estimation (whether with equal weights or not) in the experimental section, the model denoted as HQR-W (Weighted Heteroscedastic Quantile Regression), whose expression is given by \begin{equation}\label{eq:hqr_w}
\begin{split}
\widehat{q}_{\beta}(Y_{t} \vert \bm{y}_{t}) &= \widehat{\lambda}_{0}(\beta) + \widehat{\lambda}_{1}(\beta)  \widehat{y}_{t, 1} + \widehat{\lambda}_{2}(\beta)  \widehat{y}_{t, 2} + \dots \\
&\quad + \widehat{\lambda}_{M}(\beta)  \widehat{y}_{t, m} + \widehat{\lambda}_{M+1}(\beta)  s_{\widehat{y}_{t}} + \varepsilon_t(\beta), \\
\mathbb{E}\left[\varepsilon_t(\beta)\right] &= 0,
\end{split}
\end{equation}
will also be considered. This assessment is particularly important because the number of variables included in the models can differ significantly depending on whether the approximation follows (\ref{eq:cfqra}) or (\ref{eq:hqr_w}). While techniques such as L1 regularization \citep{uniejewski2021regularized} or similar approaches could be employed to address this potential issue, it is essential to first determine whether such measures are necessary at all.\\

\subsection{Width-Adaptive Conformal Inference}\label{sec:waci}

The second property we aim to achieve is maintaining the same level of confidence in the coverage level regardless of the prediction's difficulty. In other words, the coverage should remain independent of the complexity of the situation, as is the case for the true quantiles we seek to estimate \citep{feldman2021improving}. When properly estimated, the interval length serves as an indicator of prediction difficulty, reflecting both aleatoric and epistemic uncertainty. In our case, this uncertainty is captured through a quantile regression process, such as HQR. Let us denote this interval at time $T+1$ by $\widehat{C}_{\alpha, T+1}$. To ensure the desired property, we build a second interval, $\widehat{C}^c_{\alpha, T+1}$, by modifying $\widehat{C}_{\alpha, T+1}$. This adjusted interval is designed to satisfy property (\ref{eq:independence_cov_il}). From now on, the first initial interval $\widehat{C}_{\alpha, T+1}$ is called the unconformalized interval and the second one $\widehat{C}^c_{\alpha, T+1}$ the conformalized interval.\\

As we are working with time series, we will modify the ACI method to apply a different $\alpha$ as a function of time, like the original method, and also as a function of the length of the interval. Given a range of possible interval lengths from the unconformalized interval, the objective is to partition this space into smaller sub-intervals, with each sub-interval receiving a distinct correction. In other words, there is a different correction depending on the unconformalized interval length. This approach enables the differentiation of varying levels of uncertainty in the data. For instance, in Electricity Price Forecasting, two distinct regimes often emerge: one in which prices are high and the share of renewables in the energy mix is relatively small, and another in which prices are low yet uncertainty grows due to a higher proportion of renewables. The quantile regression model generating the unconformalized interval may exhibit different behaviours in these two states or may not differentiate between them when it should. In any case, it is appropriate to apply different corrections for each state. This methodology allows for the unified treatment of these states within a single framework, while also being suitable for time series data.\\

Let $\mathcal{S}$ be the conformity scores (Section \ref{sec:cqr}). Given a step $\delta \in \mathbb{R}^+$, the 1-d grid $\bm{L}$ is defined as $\bm{L} =  \left( L_{\text{min}}, L_{\text{min}} + \delta, L_{\text{min}} + 2\delta, \dots, L_{\text{max}} \right) $ whose elements belong in $ \mathbb{R}$. Let $\alpha^*$ be the objective miscoverage rate. Let's denote the element in position $i$ of a vector $\bm{v}$ by $\bm{v}\left[ i \right]$, the $p$ power of $\bm{v}$ as the $p$ power of each one of the elements of $\bm{v}$ and the absolute value of $\bm{v}$ as the absolute values of each one of the elements of $\bm{v}$. The application of WACI (Width-Adaptive Conformal Inference) over the CQR procedure looks as follows. Let $\bm{\alpha}_1 = \left(\alpha^*, \alpha^*, \dots, \alpha^* \right)$ with the same dimension as $ \bm{L}, \text{err}_1 = 0, \gamma, \sigma > 0.$

\begin{equation}\label{eq:waci}
\begin{cases}
     \bm{\alpha}_{t+1} &= \bm{\alpha}_t +\gamma \bm{w}_t(\alpha^*-\text{err}_t)\\
    \text{err}_t &= \begin{cases}
        1 & \text{if } y_t \not\in \widehat{C}_{\alpha^*, t}^c \\
        0 & \text{otherwise}
    \end{cases} \\
    \text{\textbf{dist}}_t &= \vert \bm{L} - \vert \widehat{C}_{\alpha, t} \vert \vert\\
    \bm{w}_t &= \dfrac{\text{exp}\left(\frac{ - \text{\textbf{dist}}_t^2}{2\sigma^2}\right)}{\max \left\lbrace \text{exp}\left(\frac{ - \text{\textbf{dist}}_t^2}{2\sigma^2}\right) \right\rbrace}\\
    \text{idx}_{t+1} &= \argmin_i \left\lbrace \bm{L}\left[ i \right] - \vert \widehat{C}_{\alpha, t+1} \vert \right\rbrace\\
    \Tilde{\alpha}_{t+1} &= \bm{\alpha}_{t+1} \left[ \text{idx}_{t+1} \right]\\
    \widehat{C}_{\alpha^*, t+1}^c &=  \left[ \widehat{l}_{\alpha^*_t}^c, \widehat{u}_{\alpha^*_t}^c\right] = \left[ \widehat{l}_{\alpha^*, t+1} - Q_{1-\Tilde{\alpha}_{t+1}}(\mathcal{S}), \widehat{u}_{\alpha^*, t+1} + Q_{1-\Tilde{\alpha}_{t+1}}(\mathcal{S}) \right] \\
\end{cases}
\end{equation}

\begin{figure}[h]
    \centering
    \includegraphics[width=\linewidth]{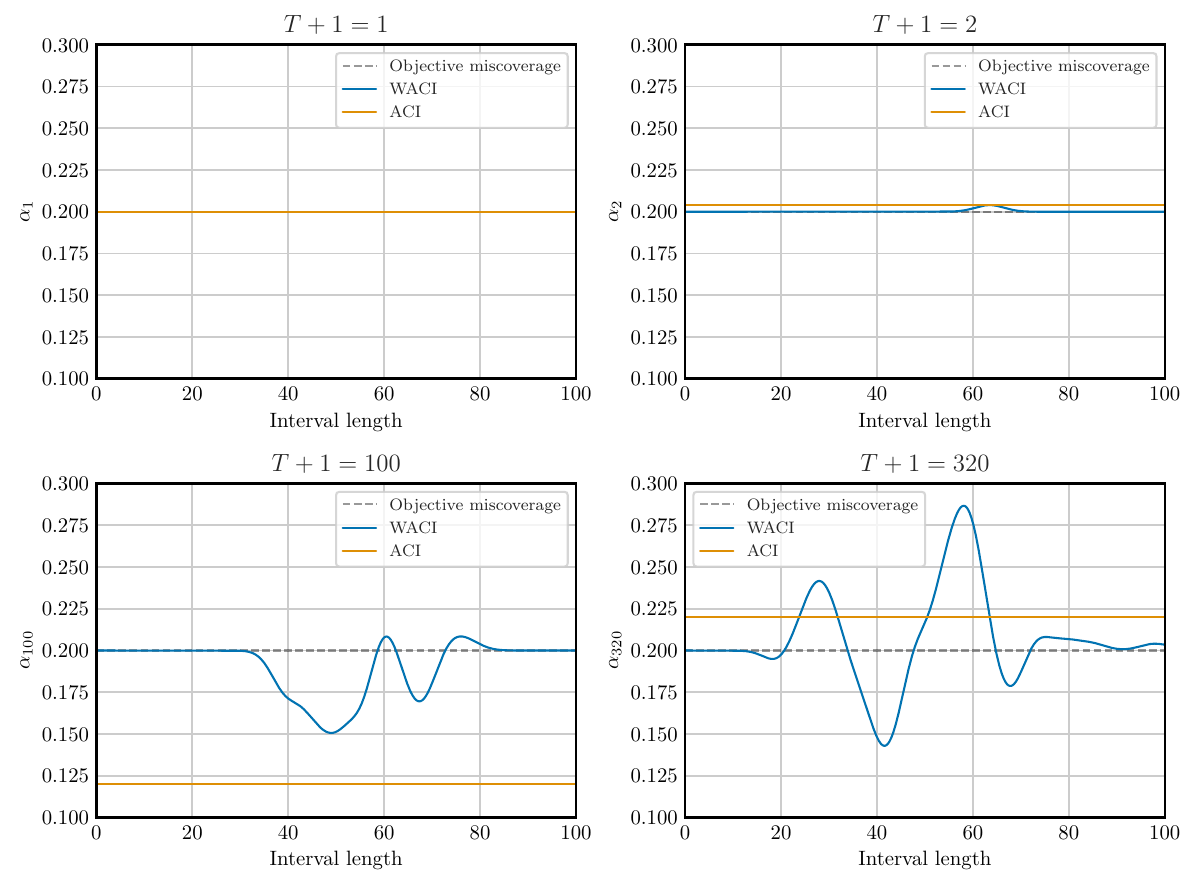}
    \caption{Evolution of $\alpha_t$ in the ACI (orange line) and WACI (blue line) methods. The $\alpha$ used in each iteration per interval length is shown.}
    \label{fig:alpha_process}
\end{figure}

The first difference that can be seen with the ACI method is that in this case there is not a single $\alpha_t$ in each iteration, but a vector $\bm{\alpha}_{t}$. This is done in order to be able to differentiate the real scalar $\Tilde{\alpha}_t$ that will actually be used in that iteration, which will depend on the length of the unconformalized interval. That is, each element of the vector is associated with a different length of the initial interval. The possible unconformalized interval lengths considered are set through the 1-d grid $\bm{L}$. Thus, $\bm{\alpha}_t\left[ i \right]$ is the $\Tilde{\alpha}$ to be used when the length of the initial interval of the observation at time $t$ is $\bm{L} \left[ i \right]$ (or $\bm{L}\left[ i \right]$ is the closest of all those considered in $\bm{L}$). The update of $\bm{\alpha}_{t}$ is done in the same way as in ACI. However, as the conformal correction is being done as a function of interval length, only the positions associated with that interval length (and close to it) are updated. To do this, the weight vector $\bm{w}_t$ is constructed through a Gaussian kernel, so a new parameter $\sigma$ related to the amplitude of the kernel effect is introduced. The difference between the ACI and WACI methods throughout iterations is shown in Figure \ref{fig:alpha_process}.\\

The upper left graph in Figure \ref{fig:alpha_process} illustrates the first iteration of both methods, where they coincide as both start with the target $\alpha$. In the next iteration (upper right graph), the methods begin to diverge, although very slightly. The ACI method adjusts $\alpha$ for all possible interval lengths, whereas the WACI method only modifies $\alpha$ for unconformalized interval lengths close to those of the previous observation. The ACI method correction will always remain constant, displaying a horizontal line since it does not differentiate between interval lengths. On the contrary, WACI exhibits variations, using significantly different alphas at ``close'' interval lengths. The lower graphs correspond to subsequent iterations. For example, in the bottom right graph, for the next iteration, if the unconformalized interval length is around 30 or 60, the correction applied is actually bigger (in the sense that $\Tilde{\alpha} > \alpha^*$) compared to the standard correction that CQR would use. However, if the unconformalized interval length is around 40, the correction is smaller. Such distinctions cannot be made by ACI.\\

\subsubsection{Weighting Scheme Considerations}

In the Algorithm (\ref{eq:waci}), an exponential decay function of distance has been chosen for the weighting scheme. However, alternative weighting schemes could be considered. For example, a scheme with fixed weights based on the position in the vector could also be used, such as one where the weights of each interval follow a geometric progression.\\

\begin{figure}[H]
    \centering
    \begin{subfigure}[t]{0.9\textwidth}
        \centering
        \hspace{-1cm}
        \begin{tikzpicture}
            \begin{axis}[
                domain=-3:3,
                samples=100,
                xlabel={$x$},
                ylabel={$y$},
                width=360px,
                height=120px,
                xtick={-3,-2,-1,0,1,2,3},
                xticklabels={20,21,22,23,24,25,26},
                axis y line=left,
                axis x line=center,
                ymin=0,
                enlarge x limits={abs=0.5},
                legend pos=north west,
                legend style={nodes={scale=0.70, transform shape}},
                grid,
                tick label style={font=\tiny},
            ]
                \addplot[seabornblue, thick] {exp(-(x-0.5)^2/2)};
                \addlegendentry{WACI Weights 1}
                \addplot[seaborngreen, thick, domain=0:1] {1};
                \addplot[seaborngreen, thick, domain=-1:0] {0.5};
                \addplot[seaborngreen, thick, domain=1:2] {0.5};
                \addplot[seaborngreen, thick, domain=-2:-1] {0.25};
                \addplot[seaborngreen, thick, domain=2:3] {0.25};
                \addplot[seaborngreen, thick, domain=-3:-2] {0.125};
                \addlegendentry{WACI Weights 2}
                \addplot[seabornblue, mark=*] coordinates {(-3,0.00219)};
                \addplot[seabornblue, mark=*] coordinates {(-2,0.04394)};
                \addplot[seabornblue, mark=*] coordinates {(-1,0.32465)};
                \addplot[seabornblue, mark=*] coordinates {(0,0.88250)};
                \addplot[seabornblue, mark=*] coordinates {(1,0.88250)};
                \addplot[seabornblue, mark=*] coordinates {(2,0.32465)};
                \addplot[seabornblue, mark=*] coordinates {(3,0.04394)};

                \addplot [seaborngreen, thick, mark=none] coordinates {(-2, 0.125) (-2, 0.25)};
                \addplot [seaborngreen, thick, mark=none] coordinates {(-1, 0.25) (-1, 0.5)};
                \addplot [seaborngreen, thick, mark=none] coordinates {(0, 0.5) (0, 1)};
                \addplot [seaborngreen, thick, mark=none] coordinates {(1, 1) (1, 0.5)};
                \addplot [seaborngreen, thick, mark=none] coordinates {(2, 0.5) (2, 0.25)};

                \addplot[seaborngreen, mark=*] coordinates {(-2.5,0.125)};
                \addplot[seaborngreen, mark=*] coordinates {(-1.5,0.25)};
                \addplot[seaborngreen, mark=*] coordinates {(-0.5,0.5)};
                \addplot[seaborngreen, mark=*] coordinates {(0.5,1)};
                \addplot[seaborngreen, mark=*] coordinates {(1.5,0.5)};
                \addplot[seaborngreen, mark=*] coordinates {(2.5,0.25)};

            \end{axis}
        \end{tikzpicture}
        \caption{}
        \label{fig:subfig_top}
    \end{subfigure}
    \hfill
    \begin{subfigure}[t]{0.9\textwidth}
        \centering
        \includegraphics[scale=0.5]{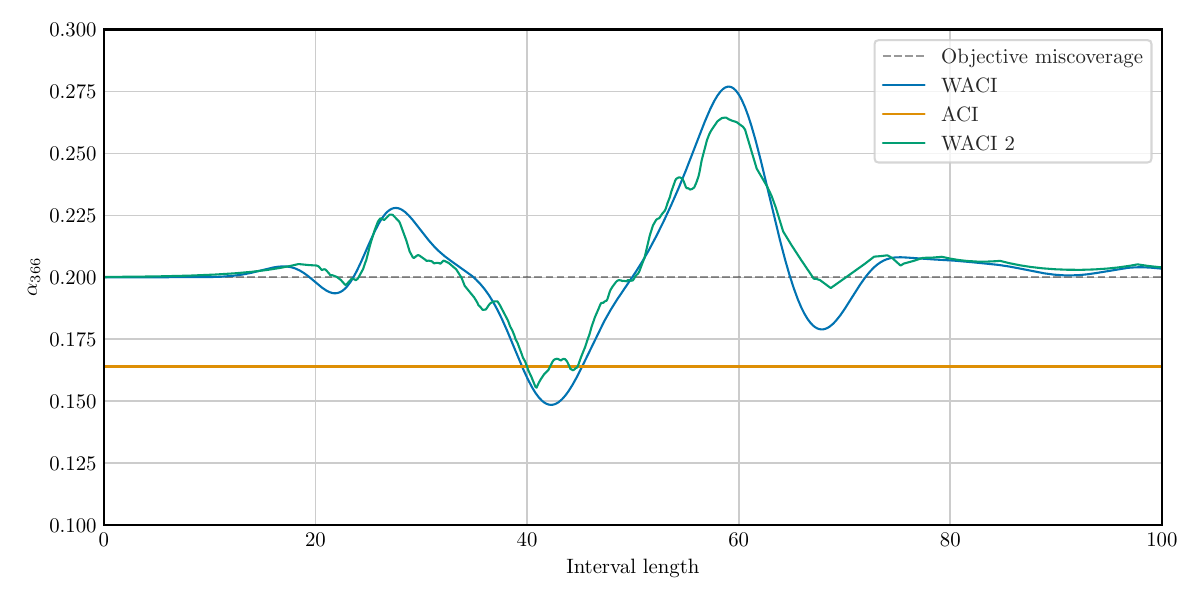}
        \caption{}
        \label{fig:subfig_bottom}
    \end{subfigure}
    \caption{(a) Comparing the different weight schemes for the WACI algorithm. The exponential decay weight is shown before scaling. (b) The behaviour of the two schemes can be very similar in practice.}
    \label{fig:gaussian_kernel}
\end{figure}

That is, let  $\bm{L}=\left(L_{\text{min}},\, L_{\text{min}} + \delta,\, L_{\text{min}} + 2\delta,\, \dots,\, L_{\text{max}} \right) \in \mathbb{R}^{*}$ and $\mathcal{L}(i_t) = \left[ \bm{L}\left[i_t\right],\bm{L}\left[i_t+1\right]\right)$ with $i_t$ the index of the interval length for the sample of instant $t$. Then,

\begin{equation}  \label{eq:pesos_2}
\bm{w}_t \left[j\right]=\lambda^{|i_t-j|}, \quad \vert C_{\alpha, t} \vert \in \mathcal{L}(i_t)
\end{equation}

Figure \ref{fig:gaussian_kernel} shows the difference between the two proposed weighting schemes. Despite their differences, by selecting the parameters $\sigma$ and $\lambda$ in a certain way, the behaviour of both is very similar.\\

If the weighting scheme (\ref{eq:pesos_2}) is considered, asymptotic conditional coverage can be proved with respect to each of the intervals considered in the grid $\bf{L}$. 

\begin{theorem}\label{thm:conditional_coverage}
    Let's assume there exists $ \nu \in \mathbb{N}$ such that $\bm{\alpha}_t \left[ i \right] \in \left[ -\nu, 1+\nu \right] \;$ for all $i=1,\ldots,n$ and $t\in\mathbb N$. Let $i\in \left\{1,...,n\right\}$ such that there is an infinite number of $t \in \mathcal{L}\left( i \right)$. If $T \longrightarrow \infty$ and the weighting scheme of (\ref{eq:pesos_2}) is considered,
    then $$ \mathbb{P}\left( y_{T+1} \in \widehat{C}^c_{\alpha^*, T+1} \, \left\vert \;  \left\vert \widehat{C}_{\alpha^*, T+1} \right\vert \right. \in \mathcal{L}\left( i \right) \right) \underset{T \longrightarrow \infty}{\longrightarrow} 1 - \alpha^*,$$ where $\alpha^*$ is the objective miscoverage rate and $\vert \widehat{C}_{\alpha^*, T+1} \vert$ is the length of the first interval produced at time step $T+1$.
\end{theorem}

\begin{proof}
The equation of the process is given by $$\bm{\alpha}_{T+1} = \bm{\alpha}_T +\gamma \bm{w}_T(\alpha^*-\text{err}_T).$$ Expanding the recursion we have $$ \bm{\alpha}_{T+1} = \bm{\alpha}_{1} + \sum_{t=1}^T \gamma \bm{w}_t \left(\alpha^* - \text{err}_t\right). $$
In particular, for each position $i$, we have $$ \bm{\alpha}_{T+1}\left[ i \right] - \bm{\alpha}_1\left[ i \right] =  \sum_{t=1}^T \gamma \bm{w}_t\left[ i \right] \left(\alpha^* - \text{err}_t\right).$$

Consider the set of indices of instants in whose iteration the length of the unconformalized interval belonged to the interval grid $j$. That is,
$$\mathcal{I}_{j} = \left\lbrace t \, : \, \left \vert \widehat{C}_{\alpha^*, t} \right \vert \in \mathcal{L}\left( j \right), \, t=1, \dots, T \right\rbrace.$$
Then, the previous expression can be decomposed based on the weight updated carried out during each iteration as
\begin{equation*}
    \bm{\alpha}_{T+1}\left[ i \right] - \bm{\alpha}_1\left[ i \right] = \sum_{j=1}^n \sum_{t \in \mathcal{I}_{j}} \gamma \lambda^{\vert i - j \vert}\left(\alpha^* - \text{err}_{t}\right) 
\end{equation*}

Denoting by $b_k=\frac{\bm{\alpha}_{T+1}\left[ k \right] - \bm{\alpha}_1\left[ k \right]}{\gamma }$ and $c_k = \sum_{t \in \mathcal{I}_{k}} \left(\alpha^* - \text{err}_{t}\right)$ for $k=1, \dots, n$; $$ b_i = c_i + \sum_{j \neq i} \lambda^{\vert i - j \vert} c_j\text{, for }i=1, \dots, n.$$

By construction, we have the following system of equations:
$$ \underbrace{\begin{pmatrix}
    b_1 \\
    b_2 \\
    \vdots \\
    b_{n} \\
\end{pmatrix}}_{\bm{b}} = 
\underbrace{\begin{pmatrix}
    1 & \lambda^{-1} & \lambda^{-2} & \dots & \lambda^{-(n-1)}\\
    \lambda^{-1} & 1 & \lambda^{-1} & \dots & \lambda^{-(n-2)}\\
    \vdots & \vdots & \vdots & \ddots & \vdots \\
    \lambda^{-(n-1)} & \lambda^{-(n-2)} & \lambda^{-(n-3)} & \dots & 1\\
\end{pmatrix}}_{\Lambda}
\underbrace{\begin{pmatrix}
    c_1 \\
    c_2 \\
    \vdots \\
    c_{n} \\
\end{pmatrix}}_{\bm{c}}
$$

The matrix $\Lambda$ is a Toeplitz matrix equivalent to the correlation matrix of a Markov-1 signal. As discussed in \cite{BRITANAK200751}, the inverse of $\Lambda$ exists (and it is known) and, therefore, $$\bm{c}= \Lambda^{-1} \bm{b}  . $$

Let $i\in \left\{1,\ldots,n\right\}$ such that $ \left \lbrace t\in \mathcal{I}_i\cap  \mathbb{N} \right \rbrace$ has an infinite number of elements and let $T_i=\# \left \lbrace t\in \mathcal{I}_i: t=1,\ldots,T \right \rbrace$. Then, as  $\Lambda^{-1}$ and $\bm{b}$ are bounded, we have
 
$$\lim_{T\rightarrow\infty} \frac 1 {T_i} \lVert \bm{c}\rVert_2 =\lim_{T\rightarrow\infty} \frac 1 {T_i} \lVert \Lambda^{-1} \bm{b}\rVert_2 =  0.$$

This implies 

$$ \lim_{T\rightarrow\infty} \frac 1 {T_i} \bm{c} = \bm{0} \Longrightarrow \lim_{T\rightarrow\infty} \frac{c_i}{T_i}  = \lim_{T\rightarrow\infty} \frac{1}{T_i} \sum_{t \in \mathcal{I}_i} \text{err}_t - \alpha^* =0,$$
which gives the result $$\mathbb{P}\left( y_{T+1} \in \widehat{C}^c_{\alpha^*, T+1} \, \left \vert \;  \left \vert \widehat{C}_{\alpha^*, T+1} \right\vert \right.\in \mathcal{L}(i) \right) \underset{T \longrightarrow \infty}{\longrightarrow} 1 - \alpha^*$$
\end{proof}

In view of Theorem \ref{thm:conditional_coverage}, asymptotic coverage conditional on the difficulty of the prediction is obtained, where that difficulty is measured by the length of the interval of the first quantile regression algorithm used. As a consequence, asymptotic marginal coverage is also achieved, as in the original ACI algorithm.\\

Although this is not exactly the condition represented by equation (\ref{eq:independence_cov_il}), coverage is still achieved depending on the complexity of the forecast. Therefore, it is important that the unconformalized interval shows the desired relationship between interval length and prediction difficulty (which it is achieved by applying HQR in the first step).\\

The only assumption made to obtain the result is that the value of $\bm{\alpha}$ is bounded for every position. Although this is not formally proven, it seems a reasonable feature of the algorithm. If the value of a certain position of $\bm{\alpha}$ exceeds 1 or falls below 0 during any iterations, it is forced to decrease or increase accordingly, thereby controlling the explosion of that value. The only scenario where no limits exists on a certain position of $\bm{\alpha}$ is when, after surpassing 1 (from above) or 0 (from below), that position is no longer frequently updated compared to others. In such cases, distant positions might continuously increase or decrease at a faster rate. This behaviour is irrational, as one would expect the algorithm to update different $\bm{\alpha}$ positions uniformly over iterations, adjusting the values both upward and downward. \\

Another implicit assumption is that $\delta$ is sufficiently small, meaning the distance between grid separator points is not excessively large. This is important because, if the grid intervals are too wide, a scenario may arise where all initial intervals fall within the same grid interval. In such cases, the procedure essentially reduces to applying ACI. When using a distance-based weighting scheme (as in equation (\ref{eq:waci})), choosing a very fine grid generally does not present a problem, aside from potentially increasing the algorithm's computational runtime. However, for position-based weighting schemes (as in equation (\ref{eq:pesos_2})), the decay of the weights along positions must reasonably reflect the distances between observations. Specifically, the weight decay should adjust based on the choice of $\delta$: if $\delta$ is small relative to the scale of the problem, the weight decay should be smoother, whereas a larger $\delta$ requires a steeper weight decay. In any case, it is recommended using as fine a grid as possible to ensure better granularity and accuracy in the results.\\

We emphasize that while HQR and WACI can be applied independently, combining them is essential to ensure that the desired properties are achieved.

\section{Computational experiments}

To evaluate the effectiveness of the proposed WACI-HQR method, the different interval construction schemes are compared. The quantile regression models include the QRA model, the HQR model, and the HQR-W model. Additionally, the conformal post-processing methods, ACI and WACI, are applied to these models. This comparison evaluates the impact of each modelling step on the interval properties, from the initial quantile regression to the final adaptive conformal approach. The weights used for WACI are given in equation (\ref{eq:waci}), with similar results observed using the weights in equation (\ref{eq:pesos_2}).

\subsection{Evaluation metrics}\label{sec:evaluation_metrics}
\paragraph{Mean empirical coverage}The mean empirical coverage is used to measure the validity property: if there are $N$ prediction intervals for observations $y_i, \, i=1, \dots, n$, the empirical coverage on those predictions is defined as $$  \frac{1}{N} \sum_{i=1}^N \mathds{1}\left( y_{i} \in  \widehat{C}_{\alpha, i} \right).$$
For an objective miscoverage rate of $\alpha$, the empirical coverage is sought to be as close to $1-\alpha$ as possible.\\

\paragraph{Mean interval length}The efficiency is usually measured through the mean or median interval length. The ACI and WACI adaptive conformal procedures are not constrained in their definition of $\alpha$, allowing for intervals that are either empty sets or cover all of $\mathbb{R}$. While empty intervals are rare, occurring only when the $\alpha$ value in the corresponding iteration exceeds 1, the case of infinite intervals is more frequent, arising when $\alpha$ is less than $0$. To address these situations, it is common practice to use the median interval length, as it is unaffected by infinite intervals. In this work, we compute the mean interval length, replacing infinite intervals with a fixed interval defined by the largest upper bound and smallest lower bound of the base model observed in the training set. This approach effectively penalizes the production of excessive infinite intervals.\\

\paragraph{Winkler score}The Winkler score \citep{winkler1972decision} is used to measure validity and efficiency together. For each time step $t$ and for a miscoverage rate of $\alpha$, it is defined as the length of the interval plus a penalty term proportional to how far the prediction is from being in the interval:
$$ W_{\alpha, t} = \begin{cases}
(\widehat{u}_{\alpha, t} - \widehat{l}_{\alpha, t}) + \frac{2}{\alpha}(\widehat{l}_{\alpha, t} - y_{t}) \quad &\text{ if } y_{t} < \widehat{l}_{\alpha, t}\\
(\widehat{u}_{\alpha, t} - \widehat{l}_{\alpha, t}) \quad &\text{ if } \widehat{l}_{\alpha, t} \leq y_{t} \leq  \widehat{u}_{\alpha, t}\\
(\widehat{u}_{\alpha, t} - \widehat{l}_{\alpha, t}) + \frac{2}{\alpha}(y_{t} - \widehat{u}_{\alpha, t}) \quad &\text{ if } y_{t} > \widehat{u}_{\alpha, t}\\
\end{cases}$$

Thus, better intervals will have smaller Winkler score. The Winkler score is actually a proper scoring rule \citep{gneiting2007strictly} and so, the mean Winkler score over every interval forecast will be also measured.\\

\paragraph{Pearson's correlation}Following \cite{feldman2021improving}, conditional coverage is assessed by computing Pearson's correlation coefficient between the interval length and the coverage indicator function. The closer to 0, the better, as the indicator function of coverage should be independent of the length when true quantiles are considered.\\

\paragraph{ILS $\lambda$ Coverage}A variation of the metric $\Delta$ILS-Coverage presented in \cite{feldman2021improving} is used to check the effectiveness of the post-processing methodologies. To evaluate whether the modifications made to the intervals produced by quantile regression algorithms are truly useful, consider a base quantile regression model with produces a prediction interval $\widehat{C}_i$ and a conformal procedure applied post hoc to this base model which produces the prediction interval $\widehat{C}_i^c$ for observation $i,\, i=1, \dots, N $. Let $\Delta_i$ denote the difference in interval lengths proposed by the two algorithms for observation $i$, defined as 
\[
\Delta_i = \left\lvert \lvert \widehat{C}^c_i \rvert - \lvert \widehat{C}_i \rvert \right\rvert.
\] 
Now consider the $\lambda \cdot 100\%$ of samples most affected by the conformal procedure, i.e., 
\[
\text{ILS} = \left\lbrace i \, : \, \Delta_i \geq q_{\lambda}\left (\left \lbrace \Delta_i \right \rbrace _{i=1}^N\right) \right\rbrace,
\]
where $q_{\lambda}$ denotes the $\lambda$-empirical quantile. The ILS $\lambda$ Coverage metric is then defined as 
\[
\text{ILS } \lambda \text{ Coverage} =  \frac{1}{\left \vert \text{ILS}\right \vert} \sum_{i \in \text{ILS}} \left \vert \mathds{1}\left( y_{i} \in  \widehat{C}^c_{i} \right) - \left(1-\alpha\right)\right\vert .
\]

The objective is to determine whether the intervals that have been modified to a greater extent indeed achieve the desired level of coverage.\footnote{The difference with \cite{feldman2021improving} is that it evaluates only the intervals that have increased in size while ignoring those that have been reduced by the conformal procedure. We propose considering both types of modifications, since reducing the size of the interval when it does not apply should also be penalized.} We consider $\lambda=0.10$.\\

\paragraph{Spearman's correlation}The Spearman's correlation coefficient between the mean absolute error of the mean prediction of the point forecasts and the interval length will be computed to assess the strength of the relationship between prediction difficulty and interval length. Since this relationship does not need to be linear, Spearman's rank correlation coefficient is preferred over the traditional linear correlation coefficient.\\

\paragraph{Standard deviation of the interval length}The standard deviation of the interval lengths generated by an algorithm is computed to assess the algorithm's ability to distinguish between varying uncertainty contexts. This metric is closely related to the previous one, as different levels of prediction difficulty are associated with corresponding variations in the interval lengths.\\

\paragraph{MCD $\lambda$}Let $1-\alpha$ denote the target coverage level, and let's assume there are $N$ observations for which a prediction interval \( \left\lbrace\widehat{C}_{\alpha, i}\right\rbrace_{i=1}^N \) has been built. To assess the property described in equation (\ref{eq:independence_cov_il}), we define the mean coverage deviation (MCD) as follows: the data is divided into $K = \frac{100}{\lambda}$ subsets \(\left\lbrace \mathcal{G}_k \right\rbrace_{k=1}^{K}\), where each subset contains approximately $\lambda$\% of the data. The subsets are defined by the empirical quantiles of \(\left\lbrace\left\vert\widehat{C}_{\alpha, i}\right\vert\right\rbrace_{i=1}^N\), such that:
\[
 \mathcal{G}_k = \left\lbrace i \,:\, q_{\frac{k-1}{K}}\left( \left\vert\widehat{C}_{\alpha, i}\right\vert\ \right) \leq \left\vert\widehat{C}_{\alpha, i}\right\vert\ < q_{\frac{k}{K}}\left( \left\vert\widehat{C}_{\alpha, i}\right\vert\ \right)\right\rbrace, \quad k = 1, \dots, K.
\]

For each subset \( \mathcal{G}_k \), the deviation between its mean empirical coverage and the objective coverage is computed
\[
D_k = \left \vert \frac{1}{\left \vert \mathcal{G}_k\right \vert} \sum_{i \in \mathcal{G}_k} \mathds{1}\left( y_{i} \in  \widehat{C}_{\alpha, i} \right) - \left(1-\alpha\right) \right \vert.
\]

Finally, the mean coverage deviation is defined as
\[
\text{MCD} = \frac{1}{K} \sum_{k=1}^{K} D_k.
\]

A smaller value of MCD indicates better alignment of empirical coverage with the target coverage, regardless of interval length. We consider $\lambda = 5$.\\

We propose two metrics that, to the best of our knowledge, have not been utilized in similar studies: Spearman's correlation coefficient and the MCD. These metrics provide valuable insights into the performance of the various methods.\\

\subsection{A synthetic example}\label{sec:synthetic_example}

A synthetic example is designed to evaluate the effectiveness of WACI and provide insights into its behaviour. Consider a time series \( y_t \) generated from a normal distribution \( N(\mu, \sigma_t) \), where the standard deviation \( \sigma_t \) alternates between two states representing different levels of uncertainty. Specifically, the process alternates between:

\begin{itemize}
    \item A high-uncertainty state where \( \sigma_t = \sigma_1 = 7 \),
    \item A low-uncertainty state where \( \sigma_t = \sigma_2 = 2 \).
\end{itemize}

The transition between these two states is governed by a probabilistic mechanism. The process begins in the high-uncertainty state (\( \sigma_t = \sigma_1 \)), and at each time step, the probability of transitioning to the other state increases incrementally by 0.0001. Once a transition occurs, the probability resets to zero, ensuring alternating states. A binary indicator variable \( \delta_t \) is used to represent the current state: \( \delta_t = 0 \) corresponds to the high-uncertainty state, while \( \delta_t = 1 \) represents the low-uncertainty state.\\

For simplicity, the mean \( \mu = 100 \) is assumed to be known. To compute the length of the interval it is simulated that a sample of size 10 is drawn from the distribution at each time step \( t \). However, instead of estimating the standard deviation from a sample, a deterministic relationship is used for \( \widehat{\sigma}_t \), introducing smooth time-dependent fluctuations that reflect transitions between overcoverage and undercoverage. This is given by:

\[
\widehat{\sigma}_t = 
\begin{cases}
    \sigma_1 + 2 \cdot \sin(0.001 \cdot t) & \text{if } \delta_t = 0, \\
    \sigma_2 + \cos(0.005 \cdot t) & \text{if } \delta_t = 1.
\end{cases}
\]
and the unconformalized interval is computed based on the following relationship:
\[
\widehat{l}_{\alpha, t} = \mu - T_{1-\frac{\alpha}{2}, 9} \cdot \widehat{\sigma_t} \cdot \sqrt{1 + \frac{1}{10}}, \quad
\widehat{u}_{\alpha, t} = \mu + T_{1-\frac{\alpha}{2}, 9} \cdot \widehat{\sigma_t} \cdot \sqrt{1 + \frac{1}{10}},
\]
where \( T_{1-\frac{\alpha}{2}, 9} \) is the critical value of the \( t \)-distribution for a two-tailed test with significance level \( \alpha \) and 9 degrees of freedom.\\

This design ensures smooth changes in coverage, making the example particularly well-suited to evaluating ACI-based conformalization methods.\\

\begin{figure}[h]
    \centering
    \includegraphics[width=\linewidth]{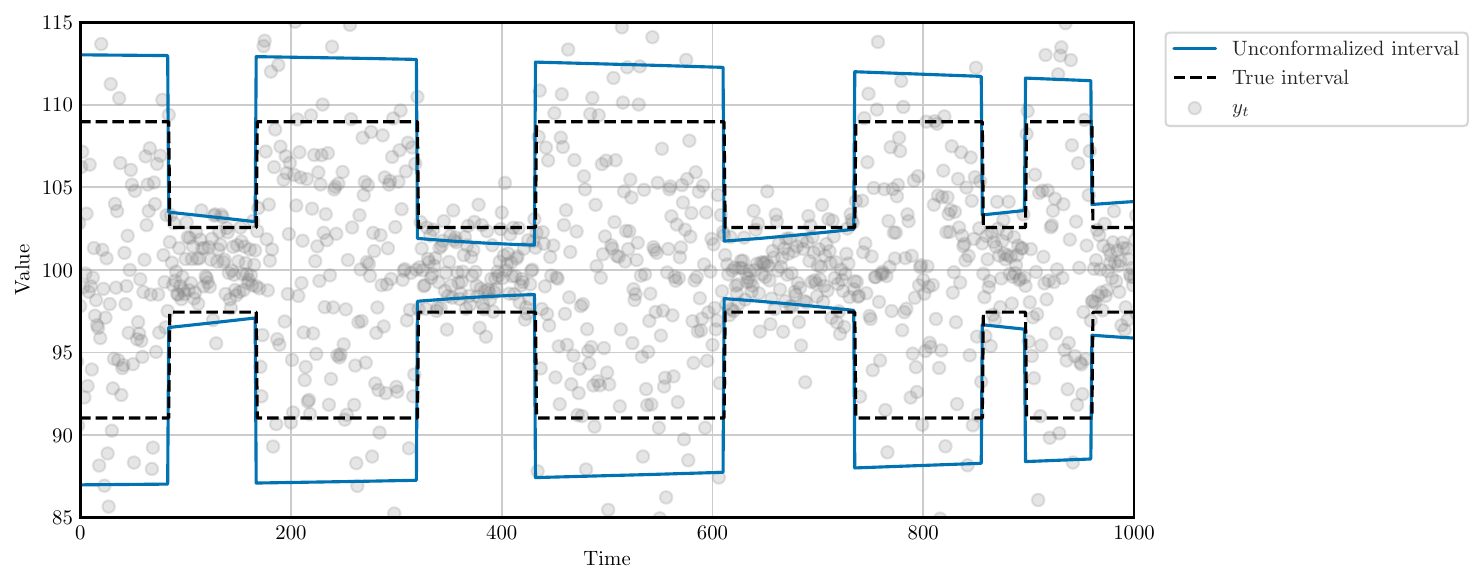}
    \caption{Simulated time series data, true intervals, and unconformalized intervals over 1000 time steps.}
    \label{fig:synthetic_example}
\end{figure}

Figure \ref{fig:synthetic_example} illustrates 1000 time steps of this process, showing the generated observations, true intervals, and unconformalized intervals.\\

The goal is to conformalize the intervals such that they remain valid and efficient across both uncertainty states, while achieving optimal performance according to the evaluation metrics introduced in Section \ref{sec:evaluation_metrics}. For this experiment, hyperparameters were fixed (\( \gamma = 0.01 \) for ACI and WACI, \( \sigma = 1 \) for WACI), and no optimization was performed. To account for the randomness inherent in the process, the experiment was repeated 100 times with different seeds, enabling standard deviation estimates for all metrics. Each run simulated a time series of length 10000 with $\alpha=0.2$.\\

\subsubsection{Results and discussion}

Tables \ref{tab:results_synhetic_1} and \ref{tab:results_synhetic_2} present the values of the different evaluation metrics previously described for the two uncertainty states separately.\\

\begin{table}[h]
\centering
\scalebox{0.75}{
\begin{tabular}{c|cccccc}
\textbf{Method} &
  \textbf{\begin{tabular}[c]{@{}c@{}}Mean Empirical\\ Coverage\end{tabular}} &
  \textbf{\begin{tabular}[c]{@{}c@{}}Average\\ Length\end{tabular}} &
  \textbf{\begin{tabular}[c]{@{}c@{}}Winkler\\ Score\end{tabular}} &
  \textbf{\begin{tabular}[c]{@{}c@{}}Pearson\\ Correlation\end{tabular}} &
  \textbf{ILS 0.10} &
  \textbf{MCD 5} \\ \hline
\textbf{Initial} & 85.13 (0.68)          & 21.14 (0.23)          & 26.05 (0.22)          & 0.25 (0.01)          & --                   & 9.17 (0.52)          \\
\textbf{ACI}     & 83.24 (0.51)          & 19.90 (0.17)          & 25.48 (0.24)          & 0.22 (0.01)          & 3.30 (0.55)          & 7.25 (0.50)          \\
\textbf{WACI}    & 81.08 (0.23) & 18.59 (0.23) & \textbf{24.89 (0.27)} & \textbf{0.15 (0.01)} & \textbf{1.47 (0.30)} & \textbf{4.35 (0.37)}
\end{tabular}
}
\caption{Mean results of 100 runs of the synthetic experiment for the high uncertainty state samples. The standard deviation of the metrics is shown in brackets. }
\label{tab:results_synhetic_1}
\end{table}

\begin{table}[h]
\centering
\scalebox{0.75}{
\begin{tabular}{c|cccccc}
\textbf{Method} &
  \textbf{\begin{tabular}[c]{@{}c@{}}Mean Empirical\\ Coverage\end{tabular}} &
  \textbf{\begin{tabular}[c]{@{}c@{}}Average\\ Length\end{tabular}} &
  \textbf{\begin{tabular}[c]{@{}c@{}}Winkler\\ Score\end{tabular}} &
  \textbf{\begin{tabular}[c]{@{}c@{}}Pearson\\ Correlation\end{tabular}} &
  \textbf{ILS 0.10} &
  \textbf{MCD 5} \\ \hline
\textbf{Initial} & 79.92 (1.19) & 5.88 (0.12) & 7.97 (0.10)          & 0.34 (0.02)          & --                   & 14.27 (0.83)         \\
\textbf{ACI}     & 76.64 (0.59) & 5.13 (0.09) & 7.57 (0.10)          & 0.25 (0.02)          & 3.34 (0.70)          & 9.87 (0.59)          \\
\textbf{WACI}    & 80.72 (0.13) & 5.35 (0.07) & \textbf{7.18 (0.09)} & \textbf{0.10 (0.01)} & \textbf{0.92 (0.22)} & \textbf{4.57 (0.36)}
\end{tabular}
}
\caption{Mean results of 100 runs of the synthetic experiment for the low uncertainty state samples. The standard deviation of the metrics is shown in brackets. }
\label{tab:results_synhetic_2}
\end{table}

\begin{figure}[h]
    \centering
    \includegraphics[width=\linewidth]{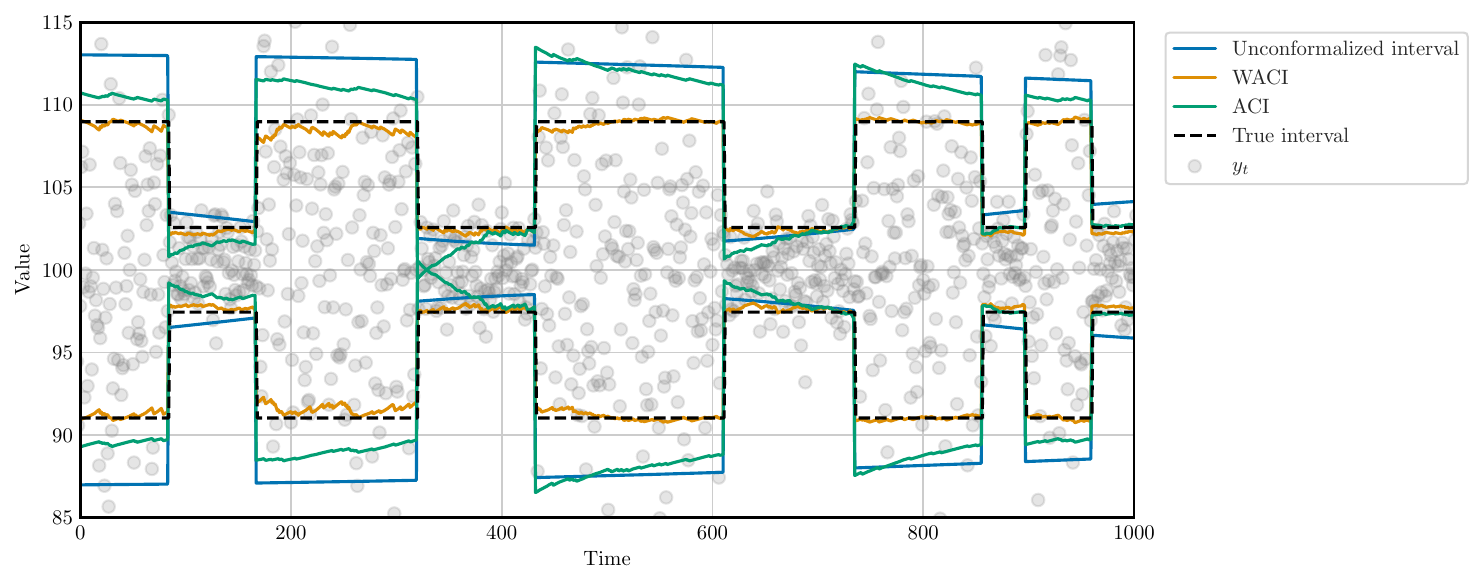}
    \caption{Prediction intervals produced by ACI (green) and WACI (orange) over the synthetic example.}
    \label{fig:predictions_synthetic}
\end{figure}

A quick comparison of the classical empirical coverage and mean interval length values reveals that WACI is the only method that effectively captures both states. This observation is further supported by the Winkler Score, which is the lowest in both cases. The other metrics provide deeper insights into why WACI is clearly the preferred choice. Notably, WACI does not exhibit the same dependence between interval length and coverage as the other two intervals, as shown in the values of the Pearson's correlation coefficient. Specifically, the large changes observed in ACI do not align closely with the desired coverage levels, unlike WACI. This difference is evident in the ILS 0.10 metric. The MCD metric highlights how WACI achieves the desired outcome by reducing the dependence between interval length and empirical coverage, resulting in intervals with superior overall characteristics. In particular, WACI allows for targeted adjustments at each uncertainty state without the need for incremental fine-tuning, as seen with ACI. This distinction is more clearly illustrated in Figure \ref{fig:predictions_synthetic}.\\

Figure \ref{fig:predictions_synthetic} illustrates how, following a change in behaviour, ACI requires several iterations to detect and adjust to the necessary modifications for approximating the true interval. This behaviour is expected since ACI relies solely on information from the previous iteration to guide its corrections. In contrast, WACI adapts immediately to the change, as it is capable of distinguishing between the two behavioural states present in the system. This is because WACI references prior iterations where the initial interval length, before conformalization, was similar.\\

It is essential to analyse the two (or more) states separately when their existence is known. Table \ref{tab:results_synhetic_3} presents the results of the evaluation metrics when all observations are considered together. At first glance, ACI appears to perform well, achieving results only slightly inferior to WACI in terms of mean empirical coverage and mean interval length. However, as shown in Tables \ref{tab:results_synhetic_1} and \ref{tab:results_synhetic_2}, this overall performance obscures significant discrepancies: in some situations, ACI exhibits over-coverage, while in others, there is clear under-coverage, leading to the observed average outcomes. Additionally, the other metrics reinforce that these discrepancies are closely tied to the relationship between interval length and coverage. The only metric for which ACI performs well is ILS 0.10, but this result warrants closer examination, particularly to identify which observations are most significantly modified by ACI.

\begin{table}[h]
\centering
\scalebox{0.75}{
\begin{tabular}{c|cccccc}
\textbf{Method} &
  \textbf{\begin{tabular}[c]{@{}c@{}}Mean Empirical\\ Coverage\end{tabular}} &
  \textbf{\begin{tabular}[c]{@{}c@{}}Average\\ Length\end{tabular}} &
  \textbf{\begin{tabular}[c]{@{}c@{}}Winkler\\ Score\end{tabular}} &
  \textbf{\begin{tabular}[c]{@{}c@{}}Pearson\\ Correlation\end{tabular}} &
  \textbf{ILS 0.10} &
  \textbf{MCD 5} \\ \hline
\textbf{Initial} & 82.51 (0.75) & 13.48 (0.44) & 19.68 (0.52)          & 0.17 (0.02)           & --                   & 11.13 (0.55)         \\
\textbf{ACI}     & 79.93 (0.02) & 12.49 (0.41) & 16.49 (0.52)          & 0.15 (0.01)           & \textbf{0.09 (0.07)} & 7.89 (0.42)          \\
\textbf{WACI}    & 80.90 (0.11) & 11.95 (0.40) & \textbf{16.01 (0.52)} & \textbf{0.04 (0.003)} & 1.17 (0.17)          & \textbf{3.68 (0.29)}
\end{tabular}
}
\caption{Mean results of 100 runs of the synthetic experiment for every observation. The standard deviation of the metrics is shown in brackets. }
\label{tab:results_synhetic_3}
\end{table}

\subsection{Electricity Price Forecasting (EPF)}\label{sec:desc_epf}
Electricity Price Forecasting serves as an ideal example for evaluating different techniques in the area of probabilistic forecasting. Producing prediction intervals in this context is of significant interest due to the intricate dynamics of electricity markets, which are increasingly influenced by stochastic factors, such as renewable energy integration. These factors directly impact the strategies of market participants and the formulation of their bids. As a result, the area of probabilistic forecasting has gained traction, with numerous studies using the electricity market as a benchmark for evaluating their methodologies. For instance, \cite{wisniewski2020application, kath2021conformal, zaffran2022adaptive}.

\begin{figure}[H]
    \centering
    \begin{subfigure}[b]{\textwidth}
        \centering
        \includegraphics[width=\textwidth]{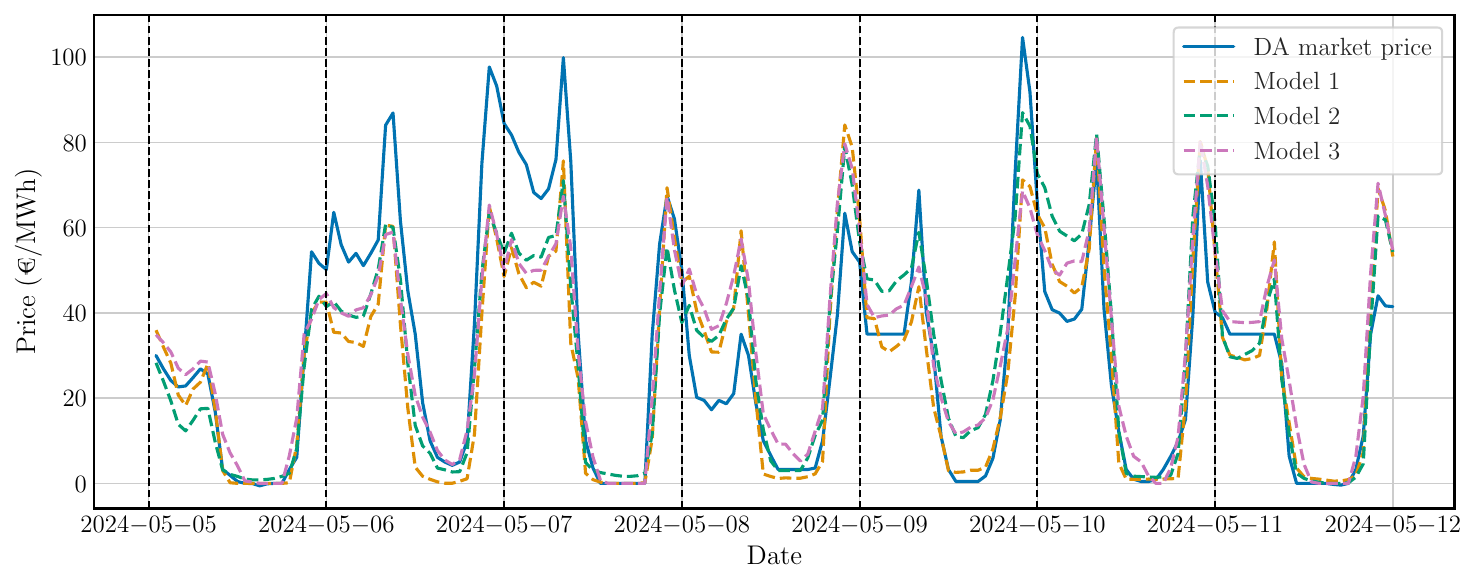}
        \caption{}
        \label{fig:spanish_market_predictions}
    \end{subfigure}
    \vspace{1em} % Add vertical spacing between the subfigures
    \begin{subfigure}[b]{\textwidth}
        \centering
        \includegraphics[width=\textwidth]{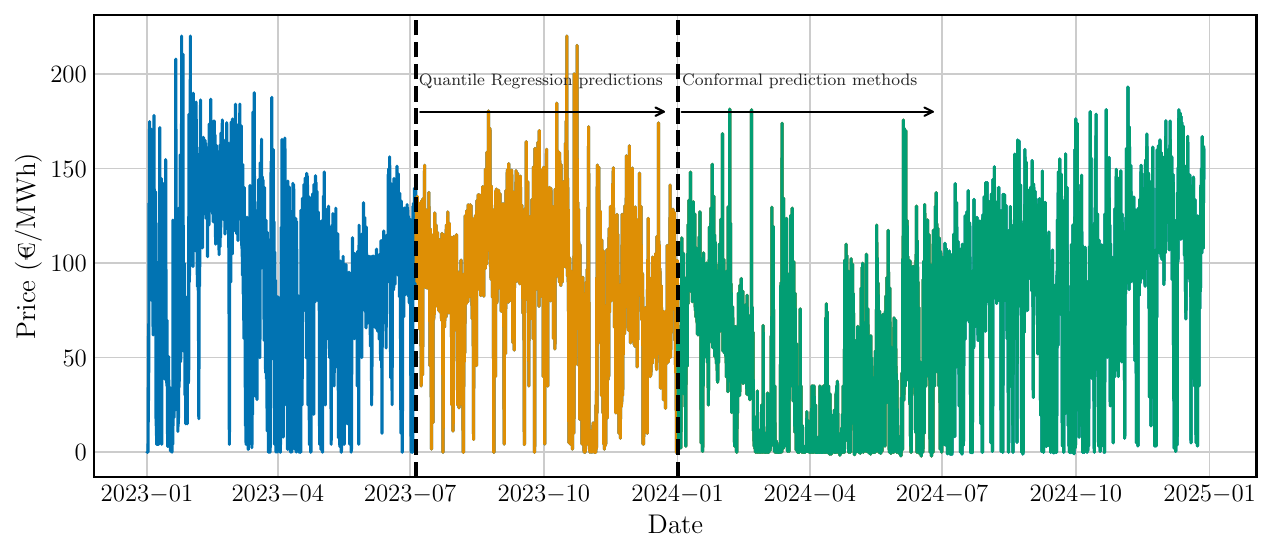}
        \caption{}
        \label{fig:spanish_market}
    \end{subfigure}
    \caption{(a) One week example of the point forecasters for the EPF example and (b) time series of the Spanish Day-Ahead market.}
    \label{fig:combined_figure}
\end{figure}

\subsubsection{Data}
Data for the Day-Ahead market price in Spain and 3 different one day-ahead point forecasters ($M=3$, Figure \ref{fig:spanish_market_predictions}) is available from 1st of January 2023 to 28th of December 2024. The period from 1st of January 2024 to 28th of December 2024 is considered as test data. This is the window where conformal methods are applied. In order for these methods to be applied, quantile regression forecasts must be available. These will be obtained from 5th of July 2023 (Figure \ref{fig:spanish_market}). This period is ideal for testing this type of models because moments of great uncertainty can be observed at the same time as very steady phases. The point forecasting models are related to the ones presented in \cite{lago2021forecasting, sebastian2023adaptive}\footnote{They can't be completely described for commercial reasons.}.\\

\subsubsection{Methodology}
When forecasting the price one day in advance, it must be predicted for the 24 hours of the next day. Thus, there are two ways of proceeding with the quantile regression: consider 24 daily series (one per hour) and a quantile regression model for each one of these series or a single hourly frequency series and therefore a single quantile regression model for all hours. As not many data is available, only the second approximation is considered. In any case, both approximations are valid and it has been shown that, for point forecasting scenarios, results are quite similar \citep{ziel2018day}.  A training rolling window of 180 days is considered. That is, at the time of predicting the day $D$, data from days $D-180$ to $D-1$ are considered to train the quantile regression models. For the day $D+1$, the window from $D-179$ to $D$ is considered for training and so on. For those days for which the training window is shorter than 180 days due to data unavailability, all available data will be used. Conformalization is always carried out individually for each hour. For both ACI and WACI $\gamma = 0.02$ is taken, which seems reasonable in view of previous studies \citep{zaffran2022adaptive} and for the WACI approach $\sigma = 3$, which also seems appropriate given the price scale. For WACI, a grid $\bm{L}$ based on a 0.1\euro/MWh step is used. With all these distinctions, 9 possible methodologies will be compared over a test period of almost one year. It is crucial to include at least one year of data to ensure the generalizability of the results \citep{lago2021forecasting}. A full year encompasses special situations such as holidays, demand seasonality, and varying meteorological conditions. This approach prevents the results from being biased or influenced by the exclusion of any particular situation, providing a comprehensive evaluation of the methods. Two possible values of $\alpha$ will be distinguished: $0.2, 0.1$, which correspond to coverage values of 80, 90\%, respectively. The calibration window size, as well as the $\gamma$ and $\sigma$ values, are hyperparameters that could be optimized to improve performance. However, in this study, such optimization has not been conducted, indicating that better results may be achievable with further tuning. To better understand the impact of $\sigma$ on the WACI methodology and the differences with ACI, an analysis can be found in the \ref{ap:behaviour_sigma}. In this example, estimations of the standard deviation for each metric have been obtained through the stationary bootstrap procedure from \cite{politis1994stationary}, considering 1000 bootstrapped samples of size 1000.\\

\subsubsection{Results and discussion}\label{sec:results}

Results for every evaluation metric are shown in Tables \ref{tab:results_epf_0.2} and 
\ref{tab:results_epf_0.1}, for $\alpha=0.20$ and $\alpha=0.10$, respectively. Metric names are displayed with their abbreviations, for better readability of the tables. On the left side of the tables, the standard metrics of mean empirical coverage, mean interval length and Winkler score are displayed. On the right side, conditional coverage metrics as well as those related to desired properties are shown.\\

\begin{table}[H]
\centering
\scalebox{0.75}{
\begin{tabular}{c|ccc|ccccc}
\textbf{Methodology} &
  \textbf{M.E.C.} &
  \textbf{M.I.L.} &
  \textbf{W.S.} &
  \textbf{P.C.} &
  \textbf{ILS 0.10} &
  \textbf{S.C.} &
  \textbf{I.L. Std} &
  \textbf{MCD 5} \\ \hline
\textbf{QRA} &
  \begin{tabular}[c]{@{}c@{}}79.68\\ (2.31)\end{tabular} &
  \begin{tabular}[c]{@{}c@{}}32.07\\ (1.65)\end{tabular} &
  \begin{tabular}[c]{@{}c@{}}49.85\\ (2.12)\end{tabular} &
  \begin{tabular}[c]{@{}c@{}}0.16\\ (0.06)\end{tabular} &
  -- &
  \begin{tabular}[c]{@{}c@{}}0.10\\ (0.10)\end{tabular} &
  \begin{tabular}[c]{@{}c@{}}9.43\\ (1.35)\end{tabular} &
  \begin{tabular}[c]{@{}c@{}}7.43\\ (1.96)\end{tabular} \\
\textbf{HQR} &
  \begin{tabular}[c]{@{}c@{}}80.24\\ (2.04)\end{tabular} &
  \begin{tabular}[c]{@{}c@{}}31.15\\ (2.06)\end{tabular} &
  \begin{tabular}[c]{@{}c@{}}47.48\\ (2.67)\end{tabular} &
  \textbf{\begin{tabular}[c]{@{}c@{}}0.04\\ (0.05)\end{tabular}} &
  -- &
  \begin{tabular}[c]{@{}c@{}}0.32\\ (0.10)\end{tabular} &
  \begin{tabular}[c]{@{}c@{}}12.54\\ (0.96)\end{tabular} &
  \begin{tabular}[c]{@{}c@{}}4.83\\ (1.39)\end{tabular} \\
\textbf{HQR-W} &
  \begin{tabular}[c]{@{}c@{}}80.78\\ (2.01)\end{tabular} &
  \begin{tabular}[c]{@{}c@{}}30.95\\ (1.90)\end{tabular} &
  \begin{tabular}[c]{@{}c@{}}47.11\\ (2.52)\end{tabular} &
  \begin{tabular}[c]{@{}c@{}}0.05\\ (0.05)\end{tabular} &
  -- &
  \begin{tabular}[c]{@{}c@{}}0.30\\ (0.10)\end{tabular} &
  \begin{tabular}[c]{@{}c@{}}12.04\\ (1.17)\end{tabular} &
  \begin{tabular}[c]{@{}c@{}}5.55\\ (1.44)\end{tabular} \\ \hline
\textbf{QRA (ACI)} &
  \begin{tabular}[c]{@{}c@{}}79.20\\ (2.12)\end{tabular} &
  \begin{tabular}[c]{@{}c@{}}32.20\\ (2.35)\end{tabular} &
  \begin{tabular}[c]{@{}c@{}}49.13\\ (2.50)\end{tabular} &
  \begin{tabular}[c]{@{}c@{}}0.20\\ (0.04)\end{tabular} &
  \begin{tabular}[c]{@{}c@{}}0.60\\ (1.28)\end{tabular} &
  \begin{tabular}[c]{@{}c@{}}0.20\\ (0.10)\end{tabular} &
  \begin{tabular}[c]{@{}c@{}}12.27\\ (1.95)\end{tabular} &
  \begin{tabular}[c]{@{}c@{}}6.91\\ (1.40)\end{tabular} \\
\textbf{HQR (ACI)} &
  \begin{tabular}[c]{@{}c@{}}79.62\\ (1.85)\end{tabular} &
  \begin{tabular}[c]{@{}c@{}}31.77\\ (2.47)\end{tabular} &
  \begin{tabular}[c]{@{}c@{}}47.13\\ (2.93)\end{tabular} &
  \begin{tabular}[c]{@{}c@{}}0.12\\ (0.03)\end{tabular} &
  \begin{tabular}[c]{@{}c@{}}0.40\\ (1.12)\end{tabular} &
  \textbf{\begin{tabular}[c]{@{}c@{}}0.35\\ (0.10)\end{tabular}} &
  \textbf{\begin{tabular}[c]{@{}c@{}}14.86\\ (1.50)\end{tabular}} &
  \begin{tabular}[c]{@{}c@{}}4.86\\ (1.32)\end{tabular} \\
\textbf{HQR-W (ACI)} &
  \begin{tabular}[c]{@{}c@{}}79.50\\ (1.90)\end{tabular} &
  \begin{tabular}[c]{@{}c@{}}31.21\\ (2.24)\end{tabular} &
  \begin{tabular}[c]{@{}c@{}}46.81\\ (2.76)\end{tabular} &
  \begin{tabular}[c]{@{}c@{}}0.14\\ (0.03)\end{tabular} &
  \begin{tabular}[c]{@{}c@{}}0.21\\ (1.16)\end{tabular} &
  \begin{tabular}[c]{@{}c@{}}0.33\\ (0.10)\end{tabular} &
  \begin{tabular}[c]{@{}c@{}}14.17\\ (1.68)\end{tabular} &
  \begin{tabular}[c]{@{}c@{}}5.69\\ (1.41)\end{tabular} \\ \hline
\textbf{QRA (WACI)} &
  \begin{tabular}[c]{@{}c@{}}80.67\\ (2.09)\end{tabular} &
  \begin{tabular}[c]{@{}c@{}}32.33\\ (1.91)\end{tabular} &
  \begin{tabular}[c]{@{}c@{}}48.99\\ (2.24)\end{tabular} &
  \begin{tabular}[c]{@{}c@{}}0.14\\ (0.04)\end{tabular} &
  \begin{tabular}[c]{@{}c@{}}1.15\\ (1.40)\end{tabular} &
  \begin{tabular}[c]{@{}c@{}}0.16\\ (0.10)\end{tabular} &
  \begin{tabular}[c]{@{}c@{}}10.03\\ (1.51)\end{tabular} &
  \begin{tabular}[c]{@{}c@{}}5.38\\ (1.63)\end{tabular} \\
\textbf{HQR (WACI)} &
  \begin{tabular}[c]{@{}c@{}}79.90\\ (2.01)\end{tabular} &
  \begin{tabular}[c]{@{}c@{}}31.16\\ (2.35)\end{tabular} &
  \begin{tabular}[c]{@{}c@{}}47.09\\ (2.77)\end{tabular} &
  \begin{tabular}[c]{@{}c@{}}0.08\\ (0.04)\end{tabular} &
  \textbf{\begin{tabular}[c]{@{}c@{}}0.08\\ (1.21)\end{tabular}} &
  \begin{tabular}[c]{@{}c@{}}0.33\\ (0.10)\end{tabular} &
  \begin{tabular}[c]{@{}c@{}}13.20\\ (1.28)\end{tabular} &
  \textbf{\begin{tabular}[c]{@{}c@{}}3.84\\ (1.25)\end{tabular}} \\
\textbf{HQR-W (WACI)} &
  \begin{tabular}[c]{@{}c@{}}80.25\\ (1.91)\end{tabular} &
  \begin{tabular}[c]{@{}c@{}}31.17\\ (2.06)\end{tabular} &
  \begin{tabular}[c]{@{}c@{}}\textbf{46.80}\\ \textbf{(2.63)}\end{tabular} &
  \begin{tabular}[c]{@{}c@{}}0.08\\ (0.04)\end{tabular} &
  \begin{tabular}[c]{@{}c@{}}0.09\\ (1.12)\end{tabular} &
  \begin{tabular}[c]{@{}c@{}}0.31\\ (0.10)\end{tabular} &
  \begin{tabular}[c]{@{}c@{}}12.52\\ (1.44)\end{tabular} &
  \begin{tabular}[c]{@{}c@{}}4.42\\ (1.24)\end{tabular}
\end{tabular}
}
\caption{Evaluation metrics for the EPF example for $\alpha=0.20$. Standard deviations of each metric are shown in brackets.}
\label{tab:results_epf_0.2}
\end{table}

\begin{table}[H]
\centering
\scalebox{0.75}{
\begin{tabular}{c|ccc|ccccc}
\textbf{Methodology} &
  \textbf{M.E.C.} &
  \textbf{M.I.L.} &
  \textbf{W.S.} &
  \textbf{P.C.} &
  \textbf{ILS 0.10} &
  \textbf{S.C.} &
  \textbf{I.L. Std} &
  \textbf{MCD 5} \\ \hline
\textbf{QRA} &
  \begin{tabular}[c]{@{}c@{}}89.10\\ (1.51)\end{tabular} &
  \begin{tabular}[c]{@{}c@{}}44.86\\ (1.99)\end{tabular} &
  \begin{tabular}[c]{@{}c@{}}63.62\\ (2.41)\end{tabular} &
  \begin{tabular}[c]{@{}c@{}}0.14\\ (0.04)\end{tabular} &
  -- &
  \begin{tabular}[c]{@{}c@{}}0.08\\ (0.10)\end{tabular} &
  \begin{tabular}[c]{@{}c@{}}11.45\\ (1.77)\end{tabular} &
  \begin{tabular}[c]{@{}c@{}}4.65\\ (1.11)\end{tabular} \\
\textbf{HQR} &
  \begin{tabular}[c]{@{}c@{}}89.58\\ (1.23)\end{tabular} &
  \begin{tabular}[c]{@{}c@{}}43.11\\ (2.63)\end{tabular} &
  \begin{tabular}[c]{@{}c@{}}59.67\\ (3.20)\end{tabular} &
  \textbf{\begin{tabular}[c]{@{}c@{}}0.06\\ (0.04)\end{tabular}} &
  -- &
  \begin{tabular}[c]{@{}c@{}}0.31\\ (0.11)\end{tabular} &
  \begin{tabular}[c]{@{}c@{}}15.79\\ (1.27)\end{tabular} &
  \begin{tabular}[c]{@{}c@{}}3.64\\ (1.27)\end{tabular} \\
\textbf{HQR-W} &
  \begin{tabular}[c]{@{}c@{}}89.61\\ (1.27)\end{tabular} &
  \begin{tabular}[c]{@{}c@{}}42.45\\ (2.20)\end{tabular} &
  \begin{tabular}[c]{@{}c@{}}59.75\\ (2.82)\end{tabular} &
  \textbf{\begin{tabular}[c]{@{}c@{}}0.06\\ (0.04)\end{tabular}} &
  -- &
  \begin{tabular}[c]{@{}c@{}}0.29\\ (0.10)\end{tabular} &
  \begin{tabular}[c]{@{}c@{}}14.36\\ (1.55)\end{tabular} &
  \begin{tabular}[c]{@{}c@{}}3.25\\ (1.00)\end{tabular} \\ \hline
\textbf{QRA (ACI)} &
  \begin{tabular}[c]{@{}c@{}}89.50\\ (1.36)\end{tabular} &
  \begin{tabular}[c]{@{}c@{}}46.36\\ (2.85)\end{tabular} &
  \begin{tabular}[c]{@{}c@{}}63.04\\ (2.81)\end{tabular} &
  \begin{tabular}[c]{@{}c@{}}0.16\\ (0.04)\end{tabular} &
  \begin{tabular}[c]{@{}c@{}}0.41\\ (1.38)\end{tabular} &
  \begin{tabular}[c]{@{}c@{}}0.18\\ (0.10)\end{tabular} &
  \begin{tabular}[c]{@{}c@{}}16.50\\ (2.29)\end{tabular} &
  \begin{tabular}[c]{@{}c@{}}4.54\\ (2.29)\end{tabular} \\
\textbf{HQR (ACI)} &
  \begin{tabular}[c]{@{}c@{}}89.84\\ (1.08)\end{tabular} &
  \begin{tabular}[c]{@{}c@{}}43.81\\ (3.14)\end{tabular} &
  \begin{tabular}[c]{@{}c@{}}59.38\\ (3.71)\end{tabular} &
  \begin{tabular}[c]{@{}c@{}}0.11\\ (0.03)\end{tabular} &
  \begin{tabular}[c]{@{}c@{}}0.08\\ (1.10)\end{tabular} &
  \textbf{\begin{tabular}[c]{@{}c@{}}0.34\\ (0.10)\end{tabular}} &
  \textbf{\begin{tabular}[c]{@{}c@{}}19.53\\ (2.00)\end{tabular}} &
  \begin{tabular}[c]{@{}c@{}}3.32\\ (0.85)\end{tabular} \\
\textbf{HQR-W (ACI)} &
  \begin{tabular}[c]{@{}c@{}}89.61\\ (1.20)\end{tabular} &
  \begin{tabular}[c]{@{}c@{}}44.04\\ (2.94)\end{tabular} &
  \begin{tabular}[c]{@{}c@{}}59.59\\ (3.28)\end{tabular} &
  \begin{tabular}[c]{@{}c@{}}0.11\\ (0.04)\end{tabular} &
  \begin{tabular}[c]{@{}c@{}}0.24\\ (1.16)\end{tabular} &
  \begin{tabular}[c]{@{}c@{}}0.31\\ (0.10)\end{tabular} &
  \begin{tabular}[c]{@{}c@{}}18.78\\ (2.34)\end{tabular} &
  \begin{tabular}[c]{@{}c@{}}3.80\\ (0.85)\end{tabular} \\ \hline
\textbf{QRA (WACI)} &
  \begin{tabular}[c]{@{}c@{}}90.31\\ (1.26)\end{tabular} &
  \begin{tabular}[c]{@{}c@{}}46.65\\ (2.11)\end{tabular} &
  \begin{tabular}[c]{@{}c@{}}62.98\\ (2.52)\end{tabular} &
  \begin{tabular}[c]{@{}c@{}}0.13\\ (0.03)\end{tabular} &
  \begin{tabular}[c]{@{}c@{}}0.38\\ (1.27)\end{tabular} &
  \begin{tabular}[c]{@{}c@{}}0.13\\ (0.10)\end{tabular} &
  \begin{tabular}[c]{@{}c@{}}12.97\\ (1.71)\end{tabular} &
  \begin{tabular}[c]{@{}c@{}}3.72\\ (1.05)\end{tabular} \\
\textbf{HQR (WACI)} &
  \begin{tabular}[c]{@{}c@{}}90.14\\ (1.21)\end{tabular} &
  \begin{tabular}[c]{@{}c@{}}43.52\\ (2.99)\end{tabular} &
  \textbf{\begin{tabular}[c]{@{}c@{}}59.35\\ (3.47)\end{tabular}} &
  \begin{tabular}[c]{@{}c@{}}0.07\\ (0.04)\end{tabular} &
  \textbf{\begin{tabular}[c]{@{}c@{}}0.06\\ (1.25)\end{tabular}} &
  \begin{tabular}[c]{@{}c@{}}0.32\\ (0.11)\end{tabular} &
  \begin{tabular}[c]{@{}c@{}}17.60\\ (1.70)\end{tabular} &
  \textbf{\begin{tabular}[c]{@{}c@{}}2.57\\ (0.91)\end{tabular}} \\
\textbf{HQR-W (WACI)} &
  \begin{tabular}[c]{@{}c@{}}90.03\\ (1.16)\end{tabular} &
  \begin{tabular}[c]{@{}c@{}}43.90\\ (2.48)\end{tabular} &
  \begin{tabular}[c]{@{}c@{}}59.65\\ (3.02)\end{tabular} &
  \begin{tabular}[c]{@{}c@{}}0.07\\ (0.04)\end{tabular} &
  \begin{tabular}[c]{@{}c@{}}0.13\\ (1.11)\end{tabular} &
  \begin{tabular}[c]{@{}c@{}}0.30\\ (0.10)\end{tabular} &
  \begin{tabular}[c]{@{}c@{}}16.90\\ (1.73)\end{tabular} &
  \begin{tabular}[c]{@{}c@{}}2.66\\ (0.81)\end{tabular}
\end{tabular}
}
\caption{Evaluation metrics for the EPF example for $\alpha=0.10$. Standard deviations of each metric are shown in brackets.}
\label{tab:results_epf_0.1}
\end{table}

\begin{figure}[h]
    \centering
    \includegraphics[width=\linewidth]{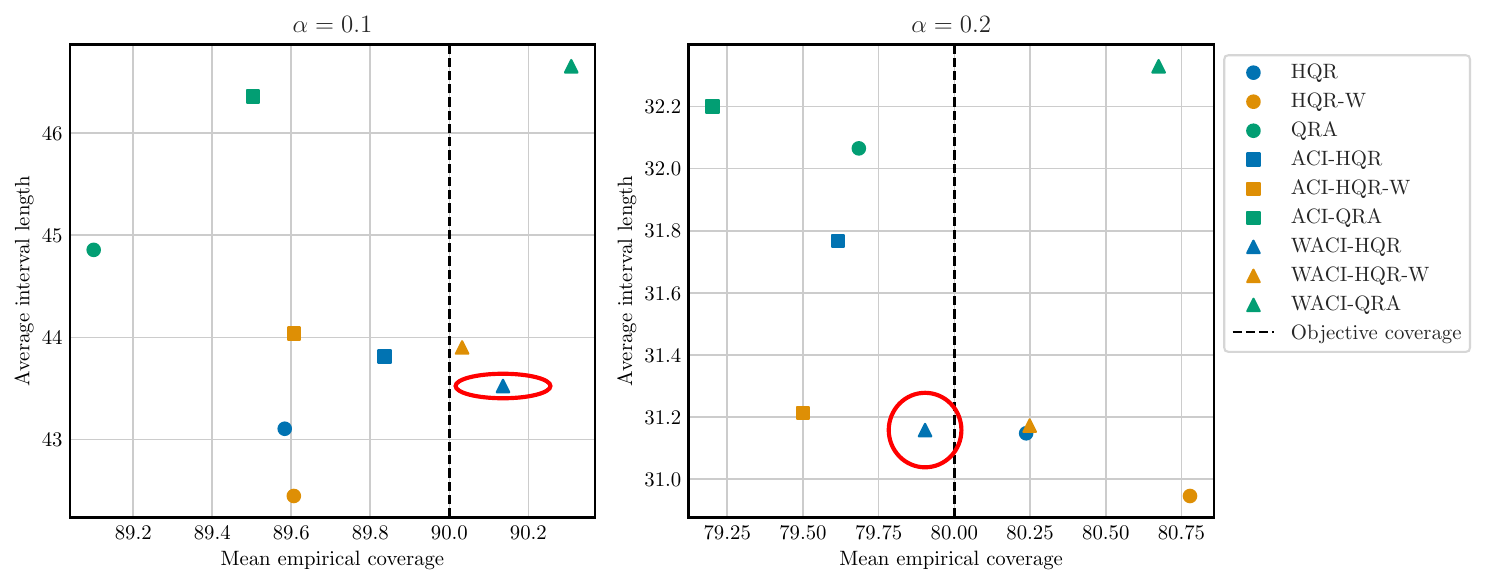}
    \caption{Mean empirical coverage vs. mean interval length for both levels of $\alpha$, 0.10 and 0.20. The colour of each point is determined by the base quantile regression model used. The shape of each point is determined by the applied conformalization. The WACI-HQR combined methodology is highlighted with a red circle.}
    \label{fig:coverage_vs_length}
\end{figure}

In order to have a better understanding of the base behaviour of the models, a plot has been made of the mean empirical coverage against the mean interval length (Figure \ref{fig:coverage_vs_length}).\\

\paragraph{Comparison between quantile regression models}At first glance, it is evident that models based on QRA perform significantly worse than those based on HQR(-W), regardless of the chosen $\alpha$ level. This is reflected in the Winkler score, where the difference is particularly pronounced, indicating poorer overall performance for QRA. Delving deeper into the metrics, several shortcomings of QRA become apparent. First, the Pearson correlation for QRA is notably higher, signalling a weaker level of conditional coverage compared to HQR(-W). Additionally, the relationship between interval length and prediction difficulty, as measured by the Spearman correlation, is considerably weaker for QRA. This suggests that QRA struggles to adjust interval lengths appropriately to reflect varying levels of uncertainty. Further evidence of this shortcoming is found in the lower standard deviation of interval lengths for QRA. This indicates an inability to differentiate between contexts of varying prediction difficulty, which complicates decision-making processes. Finally, while none of the methods are explicitly designed for this purpose, the superior approximation of quantiles achieved by HQR(-W) also benefits the MCD metric. In conclusion, QRA fails to match the quality achieved by HQR(-W) across all evaluated metrics.\\

\paragraph{Inclusion (or not) of varying weights in the HQR model}The differences between HQR and HQR-W are minimal and not statistically significant, suggesting that both models exhibit nearly identical behaviour. This observation holds true even when analysing different conformalization methods applied to the models. Based on the principle of parsimony, this suggests that HQR should be preferred. By selecting the simpler model, one can avoid potential overfitting issues that might arise from the additional complexity introduced by HQR-W.\\

\paragraph{Differences between ACI and WACI}When examining traditional metrics, ACI and WACI produce very similar results and are nearly indistinguishable, though two key observations stand out. First, WACI consistently achieves better Winkler score values when comparing the same base models and, second, produces valid prediction intervals, except for the HQR (WACI) case with $\alpha=0.20$ by a very small and reasonable margin. Pearson’s correlation reveals a clear difference: when applied to the QRA model, WACI improves conditional coverage, whereas ACI worsens it. For the HQR(-W) base model, ACI considerably degrades this property (with the coefficient being over 0.1), while WACI maintains it at levels comparable to the original model. Further insights emerge from the ILS 0.10 metric, where WACI demonstrates clear superiority, achieving minimal values and bringing interval coverage closer to the target. This robustness gives users confidence, as the methodology performs well even with significant ex-post corrections. Regarding the Spearman correlation and the standard deviation of interval lengths, WACI and ACI preserve these characteristics from the base models, a positive outcome that aligns with the intent behind using HQR. Finally, on the MCD metric, WACI once again outperforms ACI with a notable difference, demonstrating greater independence between the coverage indicator and interval length, which is the idea behind WACI's design. In conclusion, while ACI and WACI often appear similar in their outcomes, WACI consistently demonstrates superior performance across desirable properties.

\section{Conclusions and future work}\label{sec:conclusions}
In this paper we have considered the problem of obtaining prediction intervals that are built with the intention of assisting in decision making correctly. It has been discussed how the classical measures of validity and efficiency of intervals are not sufficient to be able to use these intervals in an appropriate manner. It is important that the intervals are varied in a way that this variation is directly related to the difficulty of the prediction and that the coverage does not depend on this difficulty, as it is possible to make the mistake of taking decisions with a certainty that does not correspond to the real one. Thus, one forecasting pipeline combining two innovations has been introduced. The first part consist on applying the HQR model, which focuses on the length of the intervals having the appropriate relationship with the difficulty of the prediction, and the second part involves the conformalization of the intervals produced by the quantile regression model through the WACI adaptive conformal process, which seeks uniformity of safety regardless of the difficulty. This is all considered in the context that only different forecasters of the event of interest are known, as this is a typical situation for practitioners. The different improvements provided by these models have been evaluated with two examples: one synthetic example to showcase the potential of WACI and the differences with ACI, and one related with electricity price forecasting, which is not a simple task. The results show how each of the proposed stages produces the desired results, correcting flaws in established models in the literature. Also, the inclusion of individual weights for each predictor has been shown to provide no significant improvement in the estimation of quantiles. This supports the approach of first obtaining the best possible estimate of the mean through techniques like \cite{gaillard2014second, wintenberger2017optimal, adjakossa2023kalman} and then using that estimate, along with the standard deviation of the predictors, to estimate the quantiles.\\

As futures lines of work, the HQR model uses only two explanatory variables, which are the first two (estimated) moments of the distribution to be predicted. However, moments such as skewness or kurtosis may be of interest and could contribute to the estimation of the quantiles, following the ideas set out in \cite{cornish1938moments}. Estimating these moments with so few predictors is not feasible, but if a considerable number of them is available, assessing the improvement by considering higher order moments is of interest. In addition, the variance estimation has not taken into account the individual quality of each of the provided models or the correlation between them. A correct use of this information could lead to better results, although something similar to what happens when combining point prediction models could occur, where the simplest combinations such as the mean perform remarkably well \citep{wang2023forecast}.

\newpage

\section*{Declaration of competing interest}
The authors declare that they have no known competing financial interests or personal relationships that could have appeared to
influence the work reported in this paper.

\section*{CRediT authorship contribution statement}
\textbf{Carlos Sebastián}: Writing – original draft, Visualization, Software, Methodology, Conceptualization. \textbf{Carlos E. González-Guillén}: Writing – review \& editing, Validation, Supervision, Conceptualization, Project administration, Funding acquisition. \textbf{Jesús Juan}: Writing – review \& editing, Validation, Supervision, Conceptualization, Project administration, Funding acquisition.

\section*{Data availability}
All data and code describing the algorithms presented are available to replicate the results via the link  \url{https://github.com/CCaribe9/HQR-WACI}.

\section*{Funding}
This work has been funded by grant MIG-20211033 from Centro para el Desarrollo Tecnológico Industrial, Ministerio de Universidades, and European Union-NextGenerationEU and by grant PID2023-153035NB-I00 funded by MICIU/AEI/10.13039/501100011033 and ERDF/EU. C.E.G.G. was also funded by a Re-qualification grant of Universidad Politécnica de Madrid funded by European Union-NextGenerationEU and by Ministerio de Universidades.

\newpage

\appendix
\section{Intuitive behaviour of the HQR model}\label{ap:behaviour_cfqra}
\begin{figure}[H]
    \centering
    \includegraphics[width=\textwidth]{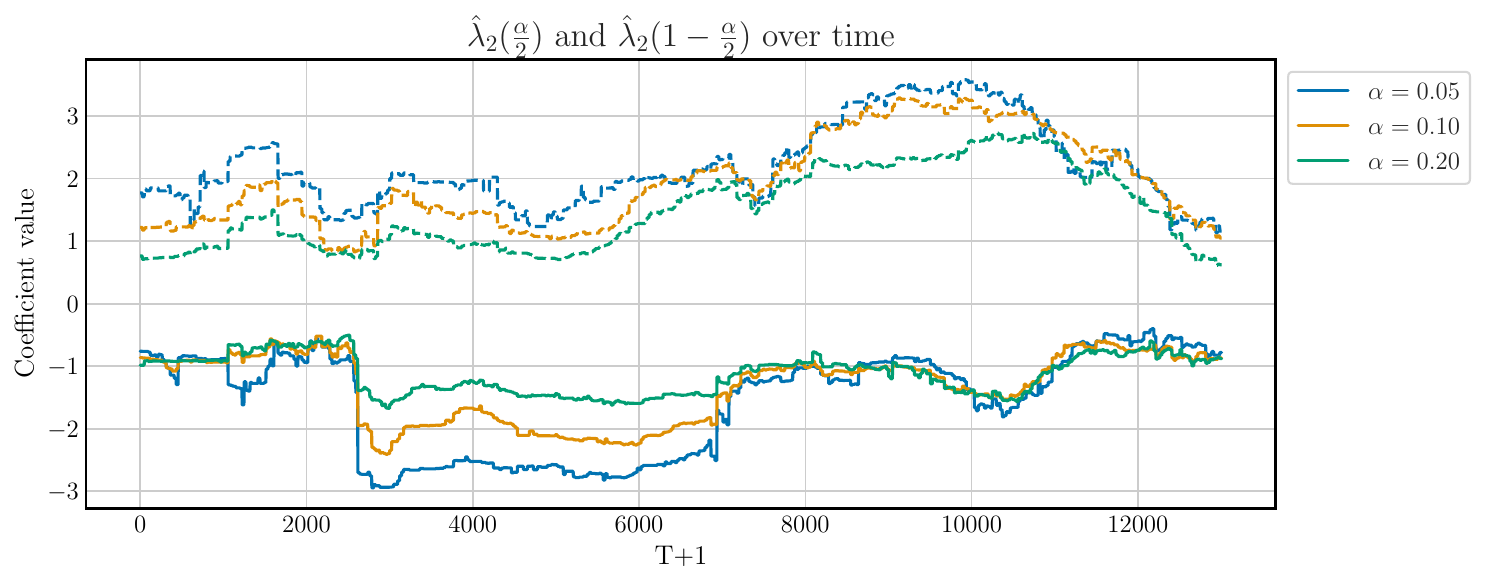}
    \caption{Value of the coefficients $\widehat{\lambda}_{2}(\frac{\alpha}{2})$ (continuous line) and $\widehat{\lambda}_{2}(1-\frac{\alpha}{2})$ (discontinuous line) for different values of $\alpha$ for the EPF example.}
    \label{fig:coefs_variation}
\end{figure}

The intuitive idea about the behaviour of the coefficients $\widehat{\lambda}_{2}(\frac{\alpha}{2})$ and $\widehat{\lambda}_{2}(1-\frac{\alpha}{2})$ is tested. These have been analysed in the example related to EPF (Section \ref{sec:desc_epf}) and the results can be seen in Figure \ref{fig:coefs_variation}. \\

As expected, the coefficients associated with the upper extremes are greater than 0, while those associated with the lower extremes are less than 0. In general, the further away the value of $\alpha$ from $0.5$, the larger the absolute value of the coefficient. In small periods of time this is not the case, which is probably related to the estimation of the other coefficients of the model. Anyway, it can be said that the intuitive idea about the expected behaviour of the model holds.\\

\section{Analysis of the impact of $\sigma$ on the WACI methodology}\label{ap:behaviour_sigma}

\begin{figure}[h]
    \centering
    \includegraphics[width=\linewidth]{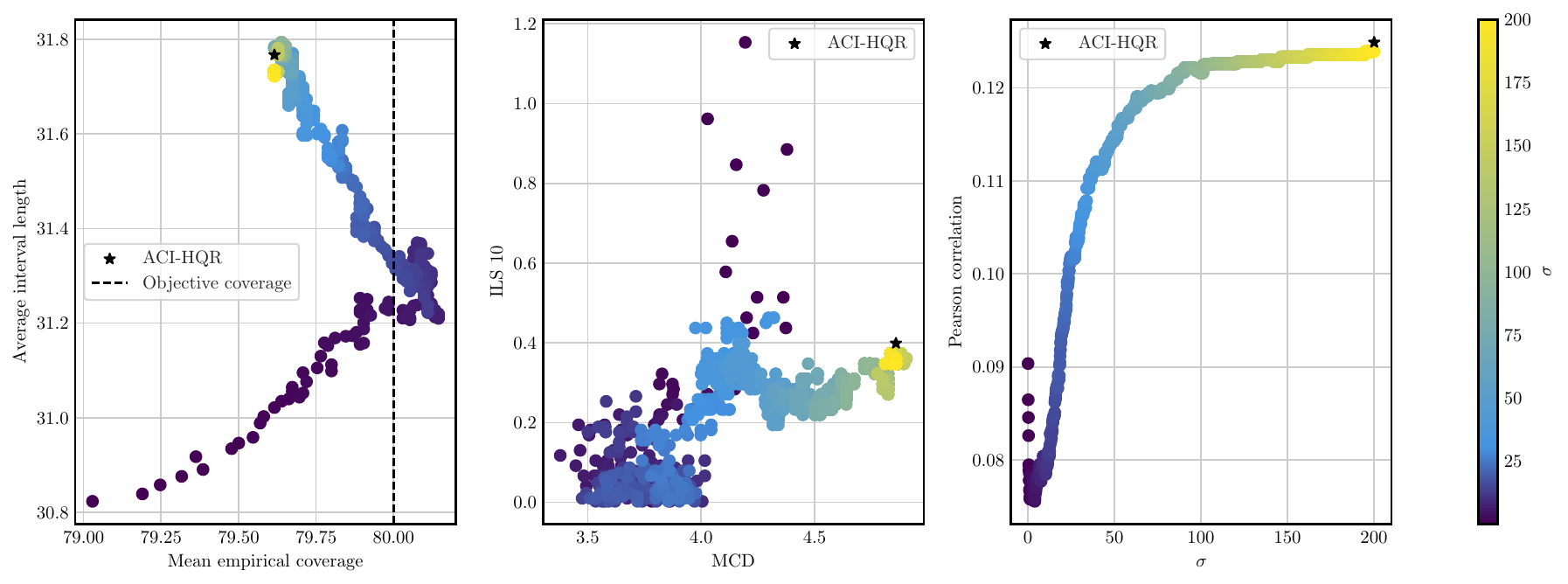}
    \caption{Mean empirical coverage vs. mean interval length, MCD vs ILS 0.10 and $\sigma$ vs Pearson's correlation plots}
    \label{fig:sigma_impact}
\end{figure}

The impact of $\sigma$ on the evaluation metrics (mean empirical coverage, mean interval length, ILS 0.10, MCD, and Pearson's correlation) has been analysed using the EPF example. For this analysis, $\gamma$ was fixed at 0.02, and the HQR model was conformalized using the WACI method with $\sigma$ values ranging from 0.1 to 200 in increments of 0.1. The only value of $\alpha$ analysed is 0.20. The results are presented in Figure \ref{fig:sigma_impact}, where the result for the ACI-HQR methodology are represented with a black star. This would be equivalent to consider $\sigma=\infty$.\\

The analysis reveals two distinct $\sigma$ scales: from 0.1 to 25 and from 25.1 to 200. Very small $\sigma$ values (less than 1) are not the best choice, as they result in negligible influence on neighbouring observations and there is not enough data to influence all interval lengths. Within the first scale, it is observed that all metrics are interrelated. Specifically, at the $\sigma$ values that yield efficient intervals, the smallest values for ILS 0.10, MCD, and Pearson's correlation are also achieved. Conversely, when the intervals become inefficient, the other metrics deteriorate significantly. Among invalid intervals, smaller $\sigma$ values outperform larger ones, achieving similar coverage with shorter intervals. Pearson's correlation also improves with smaller $\sigma$ values, except at the very smallest values, where the relationship weakens. Interestingly, there appears to be a monotonically increasing relationship between $\sigma$ and Pearson's correlation, with larger $\sigma$ values amplifying the relationship between the coverage indicator and interval length. However, large $\sigma$ values heavily penalize metrics like MCD, while ILS 0.10 is less affected.\\

\end{document}